\documentclass[12pt]{article}  % for easy procesing by JLR %\topmargin 0.2in
\usepackage{amsmath}
\usepackage{amssymb}
\usepackage{bm}            % bold math
\usepackage{cite}         %uncomment if you want to restore article instead of revtex4
\usepackage{dcolumn}       % Align table columns on decimal point
\usepackage{enumerate}
\usepackage{epsfig}
\usepackage{graphicx}
\usepackage{xcolor}
\usepackage{graphicx}

\def \bea{\begin{eqnarray}}
\def \beq{\begin{equation}}
\def \eea{\end{eqnarray}}
\def \eeq{\end{equation}}
\def \esix{E$_{\rm 6}$}

\def \s{\sqrt{2}}

\def \sx{\sqrt{6}}

\def \td{\tilde{d}}
\def \te{\tilde{e}}
\def \tiD{\tilde{D}}
\def \tL{\tilde{L}}
\def \tN{\tilde{N}}
\def \tn{\tilde{n}}
\def \tnu{\tilde{\nu}}
\def \tu{\tilde{u}}
\def \vn{\langle \tn \rangle}
\def \z2{$\mathbb{Z}_2$}

\begin{document}
\begin{titlepage}
\rightline{EFI 16-10}
%\rightline{arXiv:1609.06900}
\bigskip

\centerline{\bf SEARCHING FOR SIGNATURES OF \esix}
\medskip
\centerline{Aniket Joglekar and Jonathan L. Rosner}
% \author{\it Aniket Joglekar and Jonathan L. Rosner}
\medskip
\centerline{\it Enrico Fermi Institute and Department of Physics}
\centerline{\it University of Chicago, Chicago, IL 60637}
\bigskip

\begin{quote}
%\begin{abstract}
The grand unified group \esix~is a predictive scheme for physics beyond
the standard model (SM). It offers the possibility of extra $Z$ bosons,
new vector-like fermions, sterile neutrinos, and neutral scalars in addition
to the SM Higgs boson. Some previous discussions of these features are
updated and extended. Their relevance to present searches at the CERN Large
Hadron Collider and in patterns of neutrino masses is noted.  Addition of a
small set of scalar bosons at the TeV scale permits gauge unification near a
scale of $10^{16}$ GeV, and leads to bounds on masses of particles beyond
those in the standard model.
%\end{abstract}
\end{quote}
%\bigskip \\
\leftline{PACS categories: 12.10.Dm, 12.60.Cn, 14.70.Pw, 14.80.Ec}
%\bigskip
%\maketitle
\end{titlepage}

\centerline{\bf I.  INTRODUCTION}
\bigskip

Candidates for unification of the standard model (SM) electroweak and strong
interactions include the groups SU(5) \cite{Georgi:1974sy}, SO(10) %
\cite{Georgi:1974my, Fritzsch:1974nn}, and \esix~\cite{Gursey:1975ki}.  The
known left-handed quarks and leptons may be accommodated in three $\overline{5} + 10 +
1$ reducible representations of SU(5).  The singlets correspond to left-handed
weak isosinglet antineutrinos, needed to accommodate neutrino oscillations.
SO(10) unifies the representations in each family into three 16-dimensional
spinors, with the ``seesaw'' mechanism a popular way to understand the
smallness of neutrino masses \cite{seesaw}.  Each 27-dimensional fundamental
representation of \esix~contains not only a 16-plet spinor of SO(10), but an
SO(10) 10-plet vector and an SO(10) singlet:
\beq
27 = 16 + 10 + 1~.
\eeq
The 10-plet of SO(10) contains a $5 + \overline{5}$ of SU(5), where the
$\overline{5}$ contains an electroweak singlet color-antitriplet antiquark
$D^c$ with charge 1/3 and an electroweak lepton doublet ($L^-,L^0$).  The
pairing of $\overline{5}$ with 5 implies that the couplings of electroweak
gauge bosons to members of the SO(10) 10-plet
are purely vector-like, with no axial-vector component.  The singlet $n$ of
SO(10) has no tree-level coupling to gluons or electroweak gauge bosons,
aside from that induced by mixing with other neutral leptons.

Signatures of \esix~include extension of the Higgs sector \cite{HHG}; existence
of neutral $Z'$ gauge bosons at masses above the electroweak scale whose
decays in hadronic collisions display characteristic forward-backward
asymmetries \cite{Robinett:1982yy,Langacker:1984dc,London:1986dk,%
Hewett:1988xc}; the production of new vectorlike quarks and leptons
\cite{Rosner:2000rd,Andre:2003wc,Bjorken:2002vt}; and
manifestations of the neutral fermion $n$ through its mixing with other
neutral leptons, giving rise to signatures of ``sterile'' neutrinos
\cite{Rosner:1985hx,Rosner:2014cha,Hewett:1988xc,Nandi:1985uh,%
Mohapatra:1986bd,London:1986up,Cvetic:1992ct,Ma:1995xk,Chacko:1999aj,%
Frank:2004vg,Frank:2005rb,Dev:2012bd}.  Up to now, with the possible exception
of weak evidence for sterile neutrinos \cite{sbl1,sbl2,sbl3,%
Kopp:2013vaa, deGouvea:2013onf,Aguilar:2001ty,AguilarArevalo:2007it,%
Mention:2011rk,Mueller:2011nm,Huber:2011wv} there has been no indication
of the extra degrees of freedom entailed by the 27-plet of \esix~.

A potential change in this situation occured with claims by the ATLAS
\cite{ATLASGG,Aaboud:2016tru} and CMS \cite{CMSGG,Khachatryan:2016hje}
Collaborations at the CERN Large Hadron Collider (LHC) of a diphoton
enhancement around 750 GeV.  The accumulation of further data by both
collaborations did not confirm this effect, which now appears to have been a
statistical fluctuation \cite{ATLAS:2016eeo,ATNOGG,Khachatryan:2016yec}.
Nonetheless, great
interest was stirred in the theoretical community, leading to re-examination
of predictions of many existing schemes and invention of new ones. In the
present paper, we pursue one such avenue, updating and extending some previous
investigations of \esix.

Other recent discussions of \esix~stimulated by the initial CERN digamma
reports but with more general validity include those in \cite{Ko:2016lai,%
Chao:2016mtn,Karozas:2016hcp,Hati:2016thk,King:2016wep,Cho:2016afr,%
Athron:2016usd,Leontaris:2016wsy,Cai:2016ymq,Das:2016xuc,Das:2016vkr,%
Faraggi:2016xnm,Ashfaque:2016jha,Ashfaque:2016ydg,DelleRose:2017vvz}.
See also extensive earlier work on
\esix~in Refs.\ \cite{King:2005jy,King:2005my,King:2007uj,Howl:2008xz,%
King:2008qb,Athron:2009ue,Athron:2009bs,Athron:2009ed,Athron:2010zz,%
Hall:2010ix,Athron:2011wu,Belyaev:2012si,Athron:2012sq,Callaghan:2012rv,
Hall:2013bua,Athron:2013ipa,Athron:2012pw,Nevzorov:2012hs,Nevzorov:2013tta,%
Nevzorov:2013ixa,Athron:2015tsa,Athron:2015vxg,Faraggi:2014ica}.  Early
phenomenological analyses include ones by \cite{Franceschini:2015kwy,%
Gupta:2015zzs,Altmannshofer:2015xfo,Djouadi:2016eyy}.  For a critical review
of more than 200 papers on the initial hints of a signal see
\cite{Staub:2016dxq}, with Ref.\ \cite{Franceschini:2016gxv} proposing a
number of future experiments to pin
down related physics.  Some treatments incorporated key elements (such as
vector-like fermions) of \esix~without citing it: e.g., \cite{Bauer:2015boy,%
Dev:2015vjd,Palti:2016kew,Li:2016xcj,Jiang:2006hf}.  (The introduction of heavy
vector-like fermions avoids large contributions to the $S$ parameter of Peskin
and Takeuchi \cite{Peskin:1990zt,Peskin:1991sw,Joglekar:2012vc}.) Subgroups of
\esix~other than SU(5) $\otimes$ U(1)$_\psi \otimes$ U(1)$_\chi$, including
SU(3$)_c \otimes$ SU(3)$_L \otimes$ SU(3)$_R$ and various forms of
SU(6)$\otimes$ SU(2) \cite{Langacker:1984dc,London:1986dk,%
Robinett:1982tq,Mantilla:2016sew,Dutta:2016ach}, have been used by many authors
in variants of the present scheme.

As in Refs.\ \cite{Rosner:1985hx,Rosner:2014cha,Witten:1985xc,Candelas:1985en,%
Breit:1985ud,Dine:1985vv}, we shall assume fermion masses arise from a coupling
of two 27-plet fermions with a 27-plet scalar multiplet.  We shall label all
members of this multiplet with a tilde, without assuming that they are
supersymmetric partners of the corresponding fermions.  In particular, a scalar
state $\tn$ should exist as a counterpart to the neutral fermion $n$ described
above.  While it was tempting to associate it with the effect at 750 GeV, its
properties remain of interest even if it has not yet been observed.  Our main
focus will be to develop guidance for experimental searches
that could confirm or disprove the \esix~picture at the TeV scale. 

The \esix~symmetry is considered to be spontaneously broken
at the GUT scale, first to SO(10)$\otimes$U(1)$_\psi$, which is then broken to
SU(5)$\otimes$U(1)$_\psi\otimes$U(1)$_\chi \rightarrow$ SU(5)$\otimes$U(1)$_N$
at the same scale.  (U(1)$_N$ is that linear combination of U(1)$_\psi$ and
U(1)$_\chi$ for which the left-handed antineutrino has zero charge.  For a 
recent model incorporating it, see Ref.\ \cite{DelleRose:2017vvz}.) To achieve
this breaking, in addition to the 27-plet
scalar generation, we also must have a 78-plet of \esix. Three SM singlets in
the 78-plet --- one is a singlet under SO(10), another is a singlet under the
SU(5) contained in 45 of SO(10) and the third one is a singlet under the SM
gauge group contained in 24 of SU(5) in 45 of SO(10) --- acquire vacuum
expectation values (VEVs) of the order of the GUT scale, facilitating the
spontaneous symmetry breaking of \esix~down to the SM gauge group. Details are
described in the next section.

Our model also has a $351'$-plet scalar \cite{Slansky:1980mb} that contains
scalar diquarks, an SU(2) triplet, and an SU(3) octet with appropriate U(1)$_N$
charges (see Table~\ref{tab:351c} in Appendix A) preventing them
from contributing to proton decay. Such particles then can exist at the TeV
scale, helping to achieve unification of the SM gauge couplings and raising the
unification scale to avoid violating current bounds on proton decay processes.

In Section II we decompose a 27-plet of \esix~into its SO(10) and SU(5)
components, with U(1) subgroups arising from \esix~$\to$ SO(10) $\otimes$
U(1)$_\psi$ and SO(10) $\to$ SU(5) $\otimes$ U(1)$_\chi$ \cite{Langacker:1984%
dc,London:1986dk}. The corresponding neutral gauge bosons are denoted $Z_\psi$
and $Z_\chi$, respectively.  We adopt a linear combination of U(1)$_\psi$ and
U(1)$_\chi$ \cite{Matsuoka:1986ie,Ibanez:1986si} under which the right-hand
neutrino has zero charge, allowing it to have a large Majorana mass through a
higher-dimension operator \cite{Ma:1995xk,Callaghan:2012rv,Kinge6}. The gauge
boson coupling to this U(1)$_N$ charge will be denoted $Z_N$.
We also explain the details of symmetry breaking due to the
78-plet and unification due to low energy components of the $351'$-plet.
(Ref.\ \cite{Slansky:1981yr} contains useful group-theoretic results.)
We then enumerate \esix-invariant couplings in Sec.\ III.

Under general circumstances a $\tn$ can mix with the Higgs boson.  A general
discussion of potentials and mass matrices for (pseudo)scalars, in Sec.\ IV,
indicates conditions under which this mixing can be suppressed to acceptable
levels.  The renormalization-group evolution of gauge and Yukawa couplings,
important because of the need to avoid Landau singularities, is discussed in
Sec.\ V.  Production and decays of $\tn$ are mediated by loops of
exotic fermions in the SO(10) 10-plets belonging to the \esix~27-plet.
We discuss the decays of $\tn$ to $\gamma \gamma$, $\gamma Z$, $ZZ$, and
$W^+ W^-$ in this picture in Sec.\ VI.  Sec.\ VII treats cross sections for
$\tn$ production and observation in the $\gamma \gamma$ mode.

The properties of the heavy vector-like leptons $L$ and weak isosinglet quarks
$D$ belonging to the SO(10) 10-plet depend on their decay schemes.  We shall
adopt a \z2~symmetry \cite{Rosner:2014cha} under which SO(10) 16-plets are odd
while 10-plets and singlets are even.  This opens the possibility of stable
neutral scalars or fermions which could be dark matter candidates.  The neutral
lepton states $n$ and neutrino mixing schemes involving them, discussed in
Ref.\ \cite{Rosner:2014cha}, are updated under the assumption of an exact
\z2~symmetry in Sec.\ VIII, where we also remark briefly on the consequences of
this symmetry for dark matter.

We estimate cross sections and signatures for the heavy fermions in the SO(10)
10-plet in Sec.\ IX, and suggest diagnostics for extra neutral gauge bosons
such as $Z_N$ in Sec.\ X.  In Sec. XI, we bring together the constraints on the
various types of exotic particles to show how tightly constrained the
mass spectrum is.  We use these constraints to make future projections for the
confirmation/exclusion of the model. We conclude in Sec.\ XII.
Appendix A describes the details of $\psi$, $\chi$ and $N$ charges of the
SU(5) components of the scalar sector.  Appendix B treats details of
potentials for scalar and pseudoscalar mesons, while Appendix C is devoted to
particulars of the renormalization group evolution (RGE).
\bigskip

\centerline{\bf II. U(1) CHARGES AND MULTIPLET MEMBERS}
\bigskip

\leftline{\bf A.  Fermions}
\bigskip

Under the decomposition \esix~$\to$ SO(10) $\otimes$ U(1)$_\psi$ $\to$
SU(5) $\otimes$ U(1)$_\psi \otimes$ U(1)$_\chi$, a fermion 27-plet decomposes
as shown in Table \ref{tab:listf}.  The charge
\beq \label{eqn:qndef}
Q_N = - \frac{\sqrt{15}}{4}Q_\psi - \frac{1}{4}Q_\chi
\eeq
is that linear combination of $Q_\psi$ and $Q_\chi$ for which the left-handed
antineutrinos $N_i^c$ are neutral, allowing them to obtain large Majorana
masses via higher-dimension operators.  We use the notation of Ref.\
\cite{London:1986dk} except that in accord with common use today, exotic
isovector leptons are labeled here as $L_i$, with $i=1,2,3$ denoting family,
and isosinglet heavy quarks with charge
--1/3, called $h_i$ in the 1980s, are labeled here as $D_i$ in order to avoid
confusion with the Higgs boson.  (We will not be discussing charmed mesons,
elsewhere called $D$, in this paper.)
\bigskip

\leftline{\bf B.  Scalars}
\bigskip

Whereas we assumed three families of 27-plet fermions, we consider only a
single 27-plet of scalar bosons, whose neutral members are allowed to
obtain nonzero vacuum expectation values (VEVs).  These are listed in Table
\ref{tab:lists}.  We have adopted a tilde to denote the spin-zero partner of
the corresponding fermion first family.  This is in contrast to exceptional
supersymmetric models~\cite{King:2005jy,King:2005my,King:2007uj,Howl:2008xz,King:2008qb,Athron:2009ue,Athron:2009bs,Athron:2009ed,Athron:2010zz,Hall:2010ix,Athron:2011wu,Belyaev:2012si,Athron:2012sq,Callaghan:2012rv,Hall:2013bua,Athron:2013ipa} in which three 27-plets of fermions are
accompanied by three 27-plets of (pseudo)scalars.

We shall discuss trilinear fermion-fermion-scalar couplings systematically
in the next Section.  Meanwhile we describe the roles of VEVs of each of the
five scalars in a 27-plet.  We list the left-handed fermion pairs which form
an \esix~singlet when coupled to each scalar.  We ignore for now inter-family
mixing in quarks and leptons.  The numbers after each scalar denote values of
$(2\sx Q_\psi, 2\sqrt{10} Q_\chi, 2\sqrt{10}Q_N)$.  More details about the
effects of each VEV on neutral lepton spectra are given in Ref.\
\cite{Rosner:2014cha}.  There, a \z2~symmetry was imposed whereby 16-plet
VEVs (with \z2~quantum number --1) were suppressed in comparison with 10-plet
and singlet VEVs (with \z2~quantum number 1).  Thus,
\beq
\langle \tnu \rangle,~\langle \tN \rangle \ll \langle \tL^0_1 \rangle,~
\langle \tilde{L^0_2}^c \rangle,~\langle \tn \rangle~.
\eeq
Presently we shall compare decay schemes of 10-plet fermions in cases where
this \z2~is exact with ones where it is approximate.  A very recent work
\cite{Schwichtenberg:2017xhv} also makes use of this \z2.

% This is Table I
\begin{table}
\caption{Left-handed fermions in the 27-plet of \esix, their SO(10) and SU(5)
representations, and their U(1) charges. Subscripts $i=1,2,3$ on the fermions
denote family:  $d_{1,2,3} = (d,s,b)$; $u_{1,2,3} = (u,c,t)$; $e_{1,2,3} =
(e,\mu,\tau)$.  
\label{tab:listf}}
\begin{center}
\begin{tabular}{c c c c c c r} \hline \hline
SO(10),SU(5) & $2\sqrt{6}~Q_\psi$ & $2\sqrt{10}~Q_\chi$ & $2\sqrt{10}~Q_N$ &
 Fermion & SU(3)$_c$ & $Q$ \\ \hline 
  16,$\overline{5}$ &   1  &   3  &  --2 &  $d^c_i$  & $\overline{3}$ &  1/3  \\
           &      &      &      & $\nu_i$ &   1   &    0  \\
           &      &      &      &  $e^-_i$  &   1   &  --1  \\
   16,10   &      &  --1 &  --1 &   $u_i$   &   3   &  2/3  \\
           &      &      &      &   $d_i$   &   3   & --1/3 \\
           &      &      &      &  $u^c_i$  & $\overline{3}$ & --2/3 \\
           &      &      &      &  $e^+_i$  &   1   &   1   \\
    16,1   &      &  --5 &   0  & $N_i^c$ &   1   &   0   \\ \hline
  10,$\overline{5}$ &  --2 &  --2 &   3  & $D_i^c$ & $\overline{3}$ &  1/3  \\
           &      &      &      & $L^0_{1i}$  &   1   &   0   \\
           &      &      &      & $L^-_{1i}$  &   1   &  --1  \\
    10,5   &      &   2  &   2  &  $D_i$  &   3   & --1/3 \\
           &      &      &      & $L^+_{2i}$  &   1   &   1   \\
           &      &      &      &$L^{0c}_{2i}$ &  1   &   0   \\ \hline
    1,1    &   4  &   0  & --5  &  $n_i$  &   1   &   0   \\ \hline \hline
\end{tabular}
\end{center}
\end{table}

% This is Table II
\begin{table}
\caption{Scalar mesons in a 27-plet of \esix, their
SO(10) and SU(5) representations, and their U(1) charges.
\label{tab:lists}}
\begin{center}
\begin{tabular}{c c c c c c r} \hline \hline
SO(10),SU(5) & $2\sqrt{6}~Q_\psi$ & $2\sqrt{10}~Q_\chi$ & $2\sqrt{10}~Q_N$ &
 Meson & SU(3)$_c$ & $Q$ \\ \hline 
  16,$\overline{5}$ &   1  &   3  &  --2 &  $\td^c$  & $\overline{3}$ &  1/3  \\
           &      &      &      & $\tnu_e$ &   1   &    0  \\
           &      &      &      &  $\te^-$  &   1   &  --1  \\
   16,10   &      &  --1 &  --1 &   $\tu$   &   3   &  2/3  \\
           &      &      &      &   $\td$   &   3   & --1/3 \\
           &      &      &      &  $\tu^c$  & $\overline{3}$ & --2/3 \\
           &      &      &      &  $\te^+$  &   1   &   1   \\
    16,1   &      &  --5 &   0  & $\tN^c$ &   1   &   0   \\ \hline
  10,$\overline{5}$ &  --2 &  --2 &   3  & $\tiD^c$ & $\overline{3}$ &  1/3  \\
           &      &      &      &  $\tL^0_1$  &   1   &   0   \\
           &      &      &      &  $\tL^-_1$  &   1   &  --1  \\
    10,5   &      &   2  &   2  &  $\tiD$  &   3   & --1/3 \\
           &      &      &      &  $\tL^+_2$  &   1   &   1   \\
           &      &      & &$\tilde{L^0_2}^c$&  1   &   0   \\ \hline
    1,1    &   4  &   0  & --5  &  $\tn$  &   1   &   0   \\ \hline \hline
\end{tabular}
\end{center}
\end{table}

Neutral scalar bosons in \esix, their SM $\otimes$ U(1)$_N$-invariant couplings
to left-handed fermion pairs, and the effects of their VEVs (family indices
omitted for simplicity) are as follows:

\begin{itemize}
\item $\tnu(1,3,-2)$: $(e^+ L^-_1),~(d D^c),~(N^c L^{0c}_2)$: Mixes $e$ and
$L$, $d$ and $D$; VEV small.  Exchange can contribute to exotic fermion pair
production, e.g., in the reaction $e^+ e^- \to L^+ L^-$ or $d d^c \to
D D^c$.

\item $\tN^c(1,-5,0)$: $(d^c D),~(e^-L^+),~(\nu L^{0c}_2),~(N^c N^c)$:  Mixes 
      $e$ and $L$, $d$ and $D$; VEV small.  Provides a small Majorana mass
      contribution to $N^c$, which obtains most of its Majorana mass from a
      higher-dimension operator. Exchange can contribute to exotic fermion pair
      production, e.g., in the reaction $e^+ e^- \to L^+ L^-$ or $d d^c \to
      D D^c$.

\item $\tL^0_1(-2,-2,-3)$: $(L^{0c}_2 n),~(d d^c),~(e^+e^-)$:  Dirac
      masses for down-type quarks, charged SM leptons.  Its VEV
      is $v_d$ in the standard two-Higgs-doublet model.

\item $\tilde{L^0_2}^c(-2,2,2)$: $(L^0_1 n),~(\nu N^c), (u u^c)$:  Dirac
      masses for neutrinos (overwhelmed by seesaw), up-type quarks.  
      Its VEV is $v_u$ in the standard two-Higgs-doublet model. 

\item $\tn(4,0,-5)$: $(L^0_1 L^{0c}_2),~(L^+_2 L^-_1),~(D D^c)$: Dirac masses
for SO(10) fermionic 10-plet members.
\end{itemize}

As explained in Appendix A, the scalar 78-plet contains five
singlets under the SM gauge symmetry. The 78 (adjoint representation) of
\esix~decomposes under SO(10) as follows:
\begin{align}
78=1+45+16+\overline{16}
\end{align}
The first component, which is a singlet under SO(10), can acquire a GUT scale
VEV to break \esix $\rightarrow$SO(10)$\otimes$U(1)$_\psi$. The 45 of SO(10)
then can be decomposed into SU(5) representations as
\begin{align}
45=1+24+10+\overline{10}
\end{align}
The first component, which is a singlet under SU(5), can acquire a GUT scale
VEV to break SO(10)$\rightarrow$ SU(5)$\otimes$ U(1)$_\chi$. Now the 24 of SU(5)
contains a singlet under the SM gauge group which can also acquire a GUT scale
VEV that breaks SU(5) to the SM. There is also the breaking U(1)$_\psi\otimes$
U(1)$_\chi\rightarrow$U(1)$_N$ which is driven by a $351'$-plet VEV as
described below. Thus the symmetry surviving from the GUT scale down to the TeV
scale is SM $\otimes$ U(1)$_N$. At the TeV scale U(1)$_N$ is broken due to
the VEV of another singlet contained in the $351'$-plet that gives a $Z_N$
boson its mass. There are two other singlets under $SU(5)$ contained in the
78-plet which have nonzero U(1)$_N$ charge (see Table~\ref{tab:78} in Appendix
A). Thus in the interest of preserving U(1)$_N$
charges down to the TeV scale in order to preserve the other desired properties
of the model including the low energy diquarks that unify the gauge couplings,
we do not give VEVs to these. They have terms arising from the VEVs of the
other 78-plet singlets which give rise to masses at the GUT scale. Thus, all of
the scalar 78-plet resides at the GUT scale.

The scalar $351'$-plet decomposes under SO(10) \cite{Slansky:1981yr} as 
\begin{align}
351'=1+10+\overline{16}+54+\overline{126}+144
\end{align}  
The first component (a non-singlet of U(1)$_N$) eventually acquires a VEV at
the TeV scale that gives mass to the $Z_N$ boson. A lower bound on this VEV is
set by the maximum value of $g_N$ allowed by the unification and the
experimental search lower bounds on $Z_N'$ boson masses
\cite{Chatrchyan:2012oaa,Aad:2014cka,ATLASZP,CMSZP,Radogna,ATZN17}.
More details of this are discussed in Sec. X.  The $\overline{126}$ above, when
decomposed under SU(5), also contains a singlet which has non-zero $Q_\psi$
and $Q_\chi$ charges but no $Q_N$ charge. Thus it is the one which acquires a
GUT-scale VEV to break U(1)$_\psi\otimes$U(1)$_\chi$ to U(1)$_N$.

Finally, the unification of the SM gauge couplings at a satisfactorily high
scale requires new particles at TeV scales carrying more SU(2) and SU(3)
charges than their U(1)$_Y$ charges.  This will modify their beta functions
so as to prevent their coupling constants from unifying with U(1)$_Y$ at
energies forbidden by proton decay bounds.  Diquarks, an SU(2) triplet, and an
SU(3) color octet are suitable candidates for this. The difference in magnitude
of abelian and non-abelian charges carried by diquarks is larger than that of
any other type of fermions.  The presence of the triplet and octet helps to
adjust the beta functions without touching the U(1)$_Y$ beta function, thereby
letting the GUT scale be set above the bound
from $p\rightarrow e^+\pi^0$~\cite{Miura:2016krn}.

The new particles need to be at the multi-TeV (MTeV) scale in order to achieve
unification.  Their potential role as mediators of proton decay can be avoided
if they have U(1)$_N$ charges of magnitude $\pm 5/\sqrt{40}$ or
larger. They then cannot couple to pairs of SM fermions, whose total $Q_N$
charges are never greater in magnitude than $\pm 4/\sqrt{40}$.

The $351'$-plet contains particles which exactly possess all these properties
with $Q_N$ charges of $\pm 6/\sqrt{40}$ for diquarks and $-5/\sqrt{40}$ for the
triplet and the octet.  Details are given in Appendix A, Table~\ref{tab:351c}.
Such high charges forbid tree-level couplings with two
quarks or with a quark and a lepton which could have facilitated proton decay.
This is because all the SM particles carry $Q_N$ charges no larger than
$\pm 2/\sqrt{40}$. The SO(10) singlet with $Q_N$ charge of $10/\sqrt{40}$ is
chosen to be at MTeV scale in order to be able to break U(1)$_N$ symmetry at
that scale to give mass to the $Z_N$ boson.

\begin{figure}
\begin{center}
\includegraphics[width=0.49\textwidth]{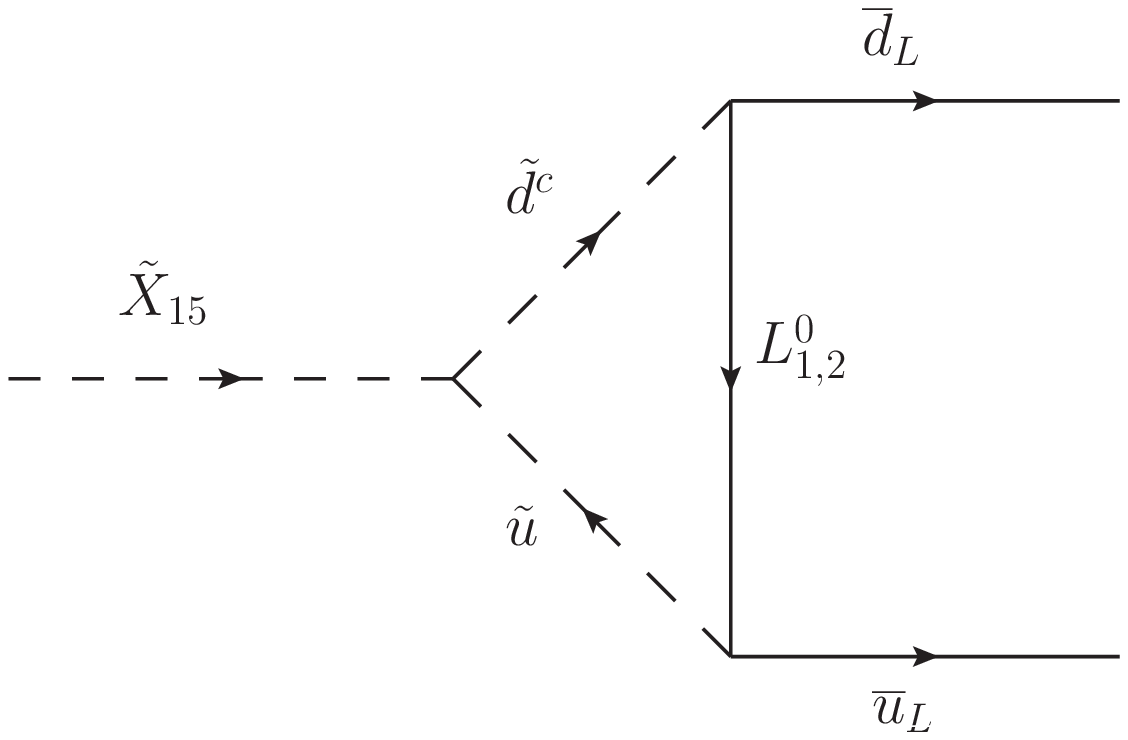}
\includegraphics[width=0.49\textwidth]{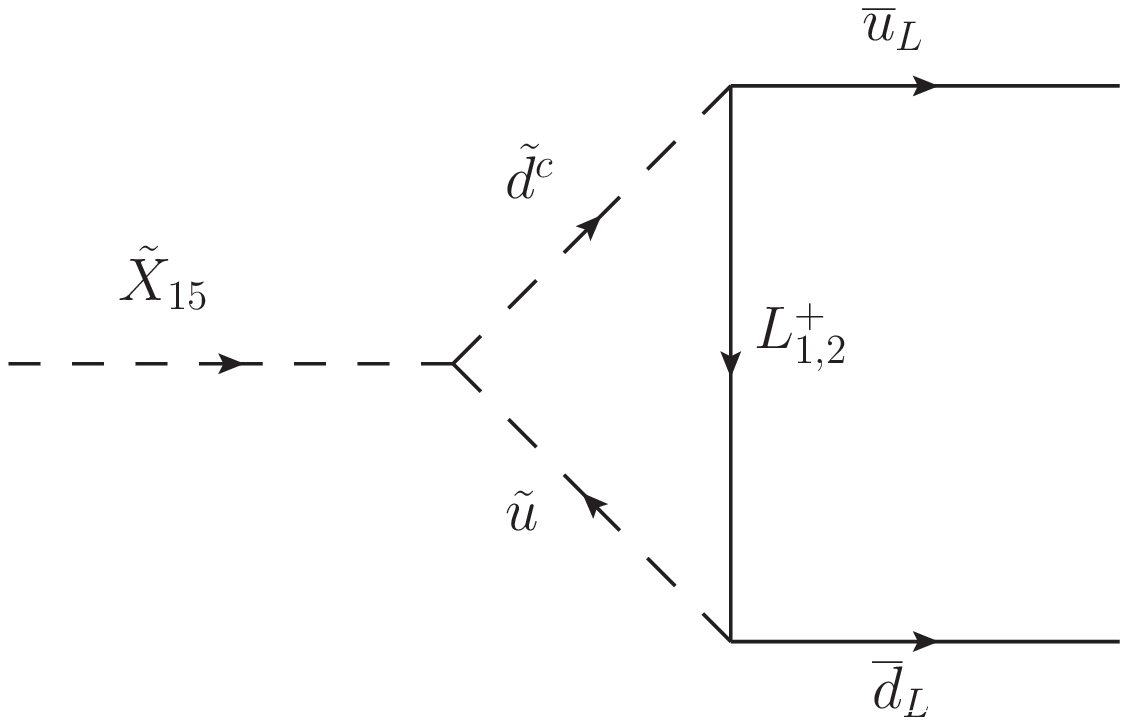}
\end{center}
\caption{$\tilde{X}_{15}$ is a diquark belonging to the 15 of SU(5) in the 54 of
SO(10) contained in the scalar $351'$ of \esix, which couples to two SM quarks 
at one loop level. The loop involves particles with GUT scale masses and
therefore is suppressed. Another multi-TeV-scale diquark, $\tilde{X}_{10}$ (not
shown) belongs to the $\overline{10}$ of SU(5) in the $\overline{16}$
of SO(10) contained in the scalar $351'$ of \esix. It does not couple to two SM
quarks at the one-loop level. For quantum numbers of $\tilde{X}_{15}$ and
$\tilde{X}_{10}$ see Table~\ref{tab:351c}.}
\label{fig:loopdiag}
\end{figure}

Loop-induced couplings of these TeV-scale states to two quarks or a quark and a
lepton are possible. But the coupling to two quarks is heavily suppressed by
the presence of the colored scalar quarks belonging to the 27-plet, as shown in
Fig.~\ref{fig:loopdiag}, which have GUT-scale masses as discussed in the next
section. Couplings to a quark and a lepton can involve exotic fermions and
scalar neutrinos belonging to 27-plets so they are not as heavily suppressed as
couplings to two quarks, but nonetheless proton decay is forbidden as it
requires both types of couplings. Other potentially dangerous operators listed
in~\cite{Arnold:2012sd} are also forbidden mainly owing to the high values of
$Q_N$ charges that these new particles carry. 

% This is Figure 1

The diquark states that belong to 15 of SU(5) contained in 54 of SO(10) and 
$\overline{10}$ of SU(5) contained in $\overline{16}$ of SO(10) (see Table
\ref{tab:351c} in Appendix A) have mass bounds due to leptoquark and dijet
searches~\cite{Khachatryan:2015qda,Sirunyan:2017yrk,Aaboud:2016qeg,Cox:2016epl,%
Khachatryan:2014lpa}.  We do not perform any explicit analysis of these, but
we avoid these bounds by assuming these diquark states to be above a few TeV.
The rich phenomenology of this sector is beyond the scope of the present work.
Details of the unification can be found in Appendix C. Thus, the $351'$-plet
plays an important role in the symmetry breaking, $Z_N$ mass, and
non-supersymmetric unification of SM gauge couplings in the present model.

It is important to note that the $351'$-plet has the same splitting problem as
the doublet-triplet splitting in the 27-plet, as some of its components reside
at the GUT scale while others reside at the TeV scale (see the mass scale
column of Table~\ref{tab:351c} in Appendix A). We assume that the model
can be fine-tuned to achieve such a large splitting owing to the large number
of parameters in the potential. The scalar 78-plet does not suffer from this
problem as it all resides at the GUT scale.

In summary, \esix~is broken to SM $\otimes$ U(1)$_N$ at the GUT scale. Then
U(1)$_N$ is broken at the TeV scale and SU(2) at the electroweak scale. At the
TeV scale or below, our model has SM gauge bosons, SM and exotic fermions; and
scalars consisting of colorless weak doublets and $\tilde{n}$ in the 27-plet,
and diquarks, SU(2) triplet, SU(3) color octet, and a U(1)$_N$-breaking SO(10)
singlet in a $351'$-plet.  This gives rise to the interesting phenomenology
discussed below and achieves one-loop unification (Appendix C).
\bigskip

\centerline{\bf III.  INVARIANT COUPLINGS}
\bigskip

We now list the {\it charged} scalars permitted to couple to left-handed
fermion pairs by invariance under the SM $\otimes$ U(1)$_N$.  Some of these
couplings will lead to proton instability unless the corresponding scalars are
very massive.  This is the famous doublet-triplet splitting problem
(\cite{Dimopoulos:1981wb}; see also \cite{Randall:1995sh,Dvali:1995hp} and
references therein).  For convenience we omit family indices on fermions.

\begin{itemize}

\item $\td^c(1,3,-2)$: $(d L^0_1),~(u L_1^-),~(N^c D),~(D^c u^c)$:  Leptoquark
and diquark.  Box diagram can contribute to flavor-changing processes and
nucleon decay.

\item $\te^-(1,3,-2)$: $(e^+ L^0_1),~(u D^c),~(N^c L_2^+)$: Mixes SM and
exotic leptons and quarks.  Exchange can contribute to $e^+ e^- \to
L^{0c}_1 L^0_2$

\item $\tu(1,-1,-1)$: $(d^c L^-_1),~(e^- D^c),~(d D),~(u^c L^{0c}_2)$:
Leptoquark and diquark.  Box diagram contributing to nucleon decay with $dd
\to L^-_1 D^c$ ($\tu$ exchange) followed by $L^-_1 h_c \to u^c \nu_c$ ($\td$
exchange)

\item $\td(1,-1,-1)$: $(d^c L^0_1),~(\nu D^c),~(u D),~(u^c L^+_2)$: Leptoquark
      and diquark. Box diagram contributing to nucleon decay with $uu \to
      L^+_2 D^c$ ($\td$ exchange) followed by $L^+_2 D^c \to d^c e^+$ ($\tu$
      exchange)

\item $\tu^c(1,-1,-1)$: $(d^c D^c),~(e^+ D),~(d L^+_2),~(u L^{0c}_2)$:
      Leptoquark and diquark.  Box diagram contributing to nucleon decay.

\item $\te^+(1,-1,-1)$: $(e^- L^0_1),~(\nu L^-),~(D u^c)$: Mixes SM and exotic
leptons and quarks.  Exchange can contribute to $e^+ e^- \to L^0_1 L^{0c}_2$.

\item $\tiD^c(-2,-2,3)$: $(d^c u^c),~(e^- u),~(\nu d),~(n D)$:  Leptoquark and
diquark; induces proton decay.

\item $\tL_1^-(-2,-2,3)$: $(d^c u),~(\nu e^+),~(L^+_2 n)$:  Charged Higgs boson.

\item $\tiD(-2,2,2)$: $(d^c N^c)~,(e^+ u^c),~(d u),~(D^c n)$:  Leptoquark and
      diquark; induces proton decay.

\item $\tL_2^+(-2,2,2)$: $(e^- N^c),~(d u^c),~(L^-_1 n)$:  Charged Higgs boson.
\end{itemize}

Many of the scalars, when exchanged, can contribute to the pair production of
exotic fermions [those in the 10 or 1 of SO(10)].  However, if these exotic
fermions have masses of order TeV or greater, they can have escaped detection
up to now. In subsequent sections we discuss the theoretical and experimental 
constraints on the Yukawa couplings and masses of these fermions. Theoretical 
constraints of perturbativity at unification scale set the upper bound on these 
masses, while the experimental searches for vector-like quarks, long-lived 
charged particles, and squarks lead to lower bounds on the masses.

More dangerous are the scalars $\tiD$ and $\tiD^c$, whose exchange
can lead, for example, to the subprocess $d u \to \bar u e^+$ and thus to
$p \to e^+ \pi^0$.  The simplest way to deal with this problem is to assume
those scalars have masses at the GUT scale.  This prevents them from being
supersymmetric partners of the vector-like quarks $D$ and $D^c$; in other words
our model does not possess TeV-scale supersymmetry. One then has to prevent the
Higgs bosons (belonging to the same SU(5) 5- or $\overline{5}$-plet as $\tiD$ or
$\tiD^c$) from acquiring large masses as well.  We shall not confront this
hierarchy problem here but an eventual solution is necessary.
\bigskip

\centerline{\bf IV.  SCALAR POTENTIAL AND BOSON MASSES}
\bigskip

As described above, the low-energy mass spectrum of the \esix~model consists
of neutral scalars and charged and neutral fermions, so signatures
for the neutral scalar $SO(10)$ singlet $\tn$ are an important feature of the
model. Such a scalar is constrained by its mixing with the SM Higgs particle in the gluon fusion channel.
In this section we investigate such a constraint as well as the nature of such
a scalar:  whether it is a real scalar or pseudoscalar in the light of a
possible future discovery at or below the TeV scale.  A $Z_N$ gauge boson is
another important TeV-scale prediction of the model, so its mass and its
relation with the $\tn$ scalars is discussed as well.
\bigskip

\leftline{\bf A.  Scalar potential}
\bigskip

The field $\tn$ is complex and may be resolved into scalar and pseudoscalar
components. A scalar can mix with the SM Higgs boson, in which
case its $\gamma \gamma$ branching fraction becomes diluted by other decay
modes such as $t\bar t$ and tree-level decays to the SM vector bosons. This is
accompanied by the reduction of the tree-level SM Higgs couplings to these
particles.  This results in constraints on the couplings of this new resonance
due to measurements of Higgs boson couplings at the LHC as well as LHC searches
for a new heavy Higgs boson decaying to a pair of SM bosons. One estimate
\cite{Falkowski:2015swt} finds the allowed mixing angle to be less than 0.1.
This problem is avoided if the $\tn$ state is taken to be a pseudoscalar,
which requires addition of another singlet or finding an alignment limit to
turn off the mixing in the scalar sector itself.  We shall explore both
possibilities.
\bigskip

The five complex scalar fields $\phi=\left(\tilde{\nu}\quad\tilde{N^c}
\quad\tilde{L_1^0}\quad\tilde{L_2^c}\quad\tilde{n}\right)$ neutral
under the SM symmetry of $SU(3)_c\otimes U(1)_{em}$ are summarized for
convenience in Table \ref{tab:neuts}.
% This is Table III
\begin{table}
	\caption{Neutral complex scalar fields belonging to the 27-plet
		or $\overline{27}$-plet of \esix.  
		\label{tab:neuts}}
	\begin{center}
	\begin{tabular}{|c c c | c c c | r |} \hline
 \multicolumn{3}{|c|}{27 member} & \multicolumn{3}{c|}{$\overline{27}$ member}
& \z2 \\
 State  & $2\sqrt{10}Q_N$ & $I_{3L}$ & State & $2\sqrt{10}Q_N$ & $I_{3L}$ & \\
\hline
$\tnu$  & --2  &  1/2  &  $\tnu^c$  &   2   & --1/2 & --1 \\
$\tN^c$  &   0  &   0   &   $\tN$    &   0   &    0  & --1 \\
$\tL_1^0$ &   3  &  1/2  & $\tL_1^{0c}$ & --3 & --1/2 &  1  \\
$\tL_2^{0c}$ &   2  & --1/2 & $\tL_2^0$ & --2 &  1/2  &  1  \\
$\tn$   & --5  &   0   &   $\tn^c$  &   5   &    0  &  1  \\ \hline \hline
	\end{tabular}
	\end{center}
\end{table}
A scalar potential can be written in terms of these fields and their conjugates
$\phi^c$.  The most general renormalizable scalar potential that obeys \esix~%
symmetry at the unification scale breaking down to $SU(3)_c\otimes SU(2)_L
\otimes U(1)_{\rm em}\otimes U(1)_N$ at the TeV scale can be written as

\begin{align}
V&=\frac{m_1^2}{2}\tilde{\nu}\tilde{\nu}^c+\frac{m_2^2}{2}\tilde{N^c}\tilde{N}
+\frac{m_3^2}{2}\tilde{L^0_1}\tilde{L^0_1}^c+\frac{m_4^2}{2}\tilde{L^0_2}
\tilde{L^0_2}^c+\frac{m_5^2}{2}\tilde{n}\tilde{n}^c
+a_1\tilde{N^c}\tilde{\nu} \tilde{L^0_2}
+a_2\tilde{N}\tilde{\nu}^c\tilde{L^0_2}^c \notag\\
&+a_3\tilde{L^0_1}\tilde{L^0_2}\tilde{n}+a_4\tilde{L^0_1}^c\tilde{L^0_2}^c
\tilde{n}^c+\frac{b_1}{2}\tilde{N^c}\tilde{\nu}\tilde{L^0_1}^c\tilde{n}^c
+\frac{b_2}{2}\tilde{N}\tilde{\nu}^c\tilde{L^0_1}\tilde{n}+\frac{b_3}{2}
\tilde{L^0_1}\tilde{L^0_1}^c\tilde{n}\tilde{n}^c+\frac{b_4}{2}\tilde{L^0_2}
\tilde{L^0_2}^c\tilde{n}\tilde{n}^c\notag\\
&+\frac{b_5}{2}\tilde{\nu}\tilde{\nu}^c\tilde{n}\tilde{n}^c+\frac{b_6}{2}
\tilde{N^c}\tilde{N}\tilde{n}\tilde{n}^c+\frac{b_7}{4}\tilde{\nu}\tilde{\nu}
\tilde{\nu}^c\tilde{\nu}^c+\frac{b_8}{2}\tilde{\nu}\tilde{N^c}\tilde{\nu}^c
\tilde{N}+\frac{b_9}{4}\tilde{N^c}\tilde{N^c}\tilde{N}\tilde{N}
+\frac{b_{10}}{2}\tilde{\nu}\tilde{\nu}^c\tilde{L^0_1}\tilde{L^0_1}^c\notag\\
&+\frac{b_{11}}{2}\tilde{\nu}\tilde{\nu}^c\tilde{L^0_2}\tilde{L^0_2}^c
+\frac{b_{12}}{2}\tilde{N^c}\tilde{N}\tilde{L^0_1}\tilde{L^0_1}^c+\frac{b_{13}}
{2}\tilde{N^c}\tilde{N}\tilde{L^0_2}\tilde{L^0_2}^c+\frac{b_{14}}{2}
\tilde{L^0_1}\tilde{L^0_1}^c\tilde{L^0_2}\tilde{L^0_2}^c+\frac{b_{15}}{4}
\tilde{L^0_1}\tilde{L^0_1}^c\tilde{L^0_1}\tilde{L^0_1}^c\notag\\
&+\frac{b_{16}}{4}\tilde{L^0_2}\tilde{L^0_2}^c\tilde{L^0_2}\tilde{L^0_2}^c
+\frac{b_{17}}{4}\tilde{n}\tilde{n}\tilde{n}^c\tilde{n}^c.
\end{align}
Quadratic terms come from the product $27 \otimes \overline{27}$; trilinear
terms come from the product $27 \otimes 27 \otimes 27$ or its charge-congugate;
quartic terms come from $27 \otimes 27 \otimes \overline{27} \otimes
\overline{27}$.  Fifteen of the seventeen quartic terms are of the form
$(\phi_a^c \phi_a)(\phi_b^c \phi_b)$.  The remaining two terms, with
coefficients $b_1$ and $b_2$, are the only additional ones found invariant when
\esix~breaks down to SO(10) and SU(5).

At the weak scale $SU(2)_L$ and $U(1)_N$ are broken spontaneously. The
spontaneous symmetry breaking of all the fields will generate the corresponding
mass matrix $\mathcal{M}$:
\begin{align}
V&=\frac{1}{2}s_i\mathcal{M}s_j,\quad\text{where}\quad\mathcal{M}
=2 \frac{\partial^2 V}{\partial s_i \partial s_j}\quad\text{and}\quad s_i=\left(\phi_k,~\phi_k^c\right).
\end{align}

We convert this basis to that of 10 real fields corresponding to real (scalar)
and imaginary (pseudoscalar) parts of the 5 complex fields $\phi_i$, as follows:
\begin{align}
\mathcal{M}_{rf}&=R^T\mathcal{M}R\quad\text{where}\quad R =\frac{1}{\sqrt{2}}
\begin{pmatrix}\mathbb{I}_{5\times5} & i\mathbb{I}_{5\times5}\\
\mathbb{I}_{5\times5} & -i\mathbb{I}_{5\times5}\end{pmatrix}
\end{align}

As described in Sec.\ II.B, \z2 symmetry leads to $\langle\tilde{\nu}\rangle,\,\langle \tilde{N}\rangle = 0$. In addition,
we assume the potential is CP even. This translates into $a_1 = a_2 \equiv a'$,
$a_3 = a_4 \equiv a$, and $b_1 = b_2 \equiv b$. These conditions result in the
separation of the scalars in the $16$ representation of $SO(10)$ from the other
three.  The elements of the corresponding $2\times2$ and $3\times3$ mass
matrices for the real parts are
$$
\mathcal{M}^s_{\tnu \tnu} = m_1^2 + b_5 \vn^2 + b_{10}v_d^2 + b_{11} v_u^2~,
$$
$$
\mathcal{M}^s_{\tnu \tN} = \mathcal{M}^s_{\tN \tnu} = 2a'v_u + b v_d \vn~,
$$
\beq
\mathcal{M}_{\tN \tN} = m_2^2 + b_6 \vn^2 + b_{12}v_d^2 + b_{13} v_u^2~;
\eeq
\smallskip

$$
\mathcal{M}^s_{\tL_1\tL_1} = m_3^2 + b_3 \vn^2 + b_{14}v_u^2 + 3 b_{15} v_d^2~,
$$
$$
\mathcal{M}^s_{\tL_1\tL_2}=\mathcal{M}^s_{\tL_2\tL_1}=2(a\vn + b_{14}v_d v_u)~,
$$
$$
\mathcal{M}^s_{\tL_2\tL_2} = m_4^2 + b_4 \vn^2 + b_{14}v_d^2 + 3 b_{16} v_u^2~,
$$
$$
\mathcal{M}^s_{\tL_1\tn} = \mathcal{M}^s_{\tn\tL_1} = 2(a v_u + b_3 v_d \vn)~,
$$
$$
\mathcal{M}^s_{\tL_2\tn}=\mathcal{M}^s_{\tn\tL_2} = 2(a v_d + b_4 v_u \vn)~,
$$
\beq \label{eqn:realscal33}
\mathcal{M}^s_{\tn\tn} = m_5^2 + b_3 v_d^2 + b_4 v_u^2 + 3b_{17} \vn^2~.
\eeq
\smallskip

The mass matrices for the pseudoscalar parts are
$$
\mathcal{M}^p_{\tnu \tnu} = m_1^2 + b_5 \vn^2 + b_{10}v_d^2 + b_{11} v_u^2~,
$$
$$
\mathcal{M}^p_{\tnu \tN} = \mathcal{M}^p_{\tN \tnu} =-2a'v_u - b v_d \vn~,
$$
\beq
\mathcal{M}^p_{\tN \tN} = m_2^2 + b_6 \vn^2 + b_{12}v_d^2 + b_{13} v_u^2~;
\eeq
\smallskip

$$
\mathcal{M}^p_{\tL_1\tL_1} = m_3^2 + b_3 \vn^2 + b_{14}v_u^2 +b_{15} v_d^2~,
$$
$$
\mathcal{M}^p_{\tL_1\tL_2} = \mathcal{M}^p_{\tL_2\tL_1} = -2a \vn~, 
$$
$$
\mathcal{M}^p_{\tL_2\tL_2} = m_4^2 + b_4 \vn^2 + b_{14}v_d^2 +b_{16} v_u^2~,
$$
$$
\mathcal{M}^p_{\tL_1\tn} = \mathcal{M}^p_{\tn\tL_1} = -2a v_u~,
$$
$$
\mathcal{M}^p_{\tL_2\tn} = \mathcal{M}^p_{\tn\tL_2} = -2a v_d~,
$$
\beq \label{eqn:pseudoscal33}
\mathcal{M}^p_{\tn\tn} = m_5^2 + b_3 v_d^2 + b_4 v_u^2 + b_{17} \vn^2~.
\eeq
where $v^2=v_u^2+v_d^2$ and $v=246\,\text{GeV}$.

The SM Higgs boson is a part of the doublet of the real scalars. Thus, in order to avoid the constraints due to SM Higgs production and decays, we would
like to turn off the mixing between the two Higgs doublets and the SO(10)
singlet.  From Eq.~(\ref{eqn:realscal33}) we see that this requires 
\begin{align}
\tan^2\beta=\frac{b_3}{b_4},\quad\text{where}\quad\tan\beta=\frac{v_u}{v_d}.
\end{align}
This also needs $b_3$ and $b_4$ to have a sign opposite to that of $a$. On the
other hand, the pseudoscalar sector mixings disappear when $a=0$. Thus the only
way to avoid doublet-singlet mixing in both real scalar and pseudoscalar
sectors is the {\em ad hoc} imposition of $a=b_3=b_4=0$. 

We will not discuss this case further as it completely decouples $\tn$ from the
SM Higgs boson, thus is not very interesting from the point of view of LHC
discovery. In the following we explore constraints on the Lagrangian parameters
in the other scenarios with small mixing. Such an analysis will also be
important in the context of supersymmetric \esix~models~\cite{King:2016wep},
where the parameter definitions are constrained by the gauge coupling constants.

We note that for $a=0$ the pseudoscalar mixings are turned off. This implies
that the Goldstone boson associated with the breaking of the U(1)$_N$ symmetry
will have to be the singlet pseudoscalar. It forces us to identify $\tn$ with 
the real singlet scalar that mixes with the SM Higgs. The fits
to the LHC Higgs coupling measurements and the heavy Higgs searches in the $WW$
and $ZZ$ channel~\cite{Falkowski:2015swt,Falkowski:2015fla} impose an upper
bound on the parameters $b_3$ and $b_4$ for a given mass spectrum. These are
relaxed in the decoupling limit, where the mass of the CP odd Higgs $(M_A)$ is
taken to several TeV.

Another possibility is to add an additional singlet complex scalar field with
the same quantum numbers as that of $\tilde{n}$. This extra singlet can be part
of an additional scalar $27$-plet that can be added to the model. Assigning a
TeV-scale vacuum value to such a singlet will decouple it from the SM Higgs.
The imaginary component of this singlet can serve as the required Goldstone
boson corresponding to the spontaneous breaking of the U(1)$_N$ symmetry, thus
making the pseudoscalar corresponding to the original singlet available at
sub-TeV masses which can be discovered at the LHC. The real scalar corresponding
to this pseudoscalar singlet can now be made much heavier than 1 TeV, even if a
sub-TeV scalar resonance is found, thus leading to its decoupling from the SM
Higgs boson and relaxation of the constraints on $b_3$ and $b_4$.

A third possibility is to turn the mixing in the scalar sector off by imposing
\begin{align}
\tan^2\beta&=\frac{b_3}{b_4}\quad\text{and}\quad a=-\frac{b_3\langle\tilde{n}\rangle}{\tan\beta}.
\end{align}
In the event of a sub-TeV scalar discovery, this case still allows for the
required pseudoscalar Goldstone boson associated with the U(1)$_N$ breaking
without the addition of an extra singlet.  Additional constraints on $b_3$
and $b_4$ will originate from the constraints on $\tan\beta$ and that on $a$
due to the fact that the global minimum (or local up to metastability) of the
complete scalar potential needs to be that corresponding to the SM electroweak
symmetry-breaking minimum.
\bigskip

\leftline{\bf B. Gauge boson masses}
\bigskip

The covariant derivative for this model is given by
\begin{align}
D_\mu&=\partial_\mu-ig\tau^aW_\mu^a-ig'YB_\mu-ig_NY_NB'_\mu
\end{align}
The scalar kinetic term in the Lagrangian that leads to the gauge boson mass
matrix is
\begin{align}
\mathcal{L}\supset \left(D^\mu \tilde{L_1}\right)^\dagger \left(D_\mu
\tilde{L_1}\right) +\left(D^\mu \tilde{L_2}\right)^\dagger \left(D_\mu
\tilde{L_2}\right)+\left(D^\mu \tilde{n}\right)^\dagger \left(D_\mu
\tilde{n}\right)
\end{align}

The charged gauge boson sector in this model remains identical to the standard
model. The neutral boson sector has an additional massive $Z_N$ boson due to
spontaneous breaking of the U(1)$_N$ symmetry. The mass matrix is as follows:
\begin{align}
\mathcal{M}&=\begin{pmatrix}
(g^2v^2)/4 & (gg'v^2)/4 & [gg_N(y_{N_1}v_d^2 +y_{N_2}v_u^2)]/2 \\
(gg'v^2)/4 & ({g'}^2 v^2)/4 & [g'g_N(y_{N_1}v_d^2 +y_{N_2}v_u^2)]/2 \\
\frac{gg_N(y_{N_1}v_d^2 + y_{N_2}v_u^2)}{2} &
\frac{g'g_N(y_{N_1}v_d^2 +y_{N_2}v_u^2)}{2} &
(g_N)^2\left(y_{N_1}^2v_d^2+y_{N_2}^2v_u^2+y_{N_s}^2 \vn^2 \right) \\
\end{pmatrix}
\end{align}
where $y_{N_1}$, $y_{N_2}$ and $y_{N_s}$ are $Q_N$ charges for the exotic
lepton doublets and SO(10) singlet, respectively.

Current lower bounds on $M(Z_N)$ are $\sim 2.5$ TeV based on 7 and 8 TeV data
\cite{Chatrchyan:2012oaa,Aad:2014cka}, and about 1 TeV higher based on 13 TeV
data up to mid-July 2016 \cite{ATLASZP,CMSZP}.  (A lower bound on $M(Z_N)$
quoted at the March 2017 Moriond Electroweak meeting, based on about 1/3 of the
13 TeV sample, was 3.41 TeV \cite{Radogna}; a recent ATLAS lower bound based on
36.1 fb$^{-1}$ is 3.8 TeV \cite{ATZN17}.)
As noted in~\cite{King:2016wep} the natural value of $g_N$ is less than one.
Another issue is the mixing of the new $Z_N$ boson with the SM neutral gauge
bosons. We need the photon to be massless; the mass of the $Z$ boson is measured
with $\sim 0.001\%$ precision. Thus the mixing needs to be small enough in
order to obey these bounds.  This implies that $g_N$ needs to be much smaller
than one. 

To satisfy the experimental lower bound on $M(Z_N)$, we can either
have $\left<\tn\right>$ of order several TeV, which in turn will impose lower  
bounds upon the mass of the singlet scalar due to requirement of the vacuum
stability as explained in Sec.~V, or we can add another singlet charged
under U(1)$_N$. This can be the SO(10) singlet of the $351'$-plet of scalars.
Let's denote it by S. As described in the previous section, such an addition
also helps with identifying the LHC-detectable $\tn$ candidate as a
pseudoscalar without going to the alignment limit in the scalar sector or {\em
ad hoc} imposing $a = b_3 = b_4 = 0$. 

Such a singlet will add a term $(g_N)^2 y_{N_s}^2\langle \tilde{S}\rangle^2$ to the diagonal element corresponding to the $Z_N$ mass.
The gauge coupling evolution shown in Fig.~\ref{fig:gauge} [Appendix C] gives
the value of $\alpha_N^{-1}$ at the experimental lower bound of the $Z_N$ boson
mass to be $\sim 86$. This means
\begin{align}
g_N^2=\frac{4\pi}{86}\approx 0.146
\end{align}
at that lower bound. This singlet has a $Q_N$ charge of $10/\sqrt{40}$. Thus,
a lower bound on the $Z_N$ mass of $\sim 4$ TeV~\cite{Chatrchyan:2012oaa,Aad:2014cka,ATLASZP,CMSZP,Radogna,ATZN17} (details in Sec.\ X) puts a lower
bound on the VEV of such a singlet:
\begin{align}
\langle S\rangle>\sqrt{\frac{16}{(0.146)(100/40)}}\sim 6.6\,\,\text{TeV}
\end{align}
This new vacuum value will decouple the $Z_N$ from the SM neutral gauge bosons,
solving both the problem of mixing with the $Z$ boson and the lower mass bound
on $Z_N$.
\bigskip

\centerline{\bf V. RENORMALIZATION GROUP EVOLUTION}
\bigskip

It is necessary to calculate the renormalization group evolution (RGE) of both
the gauge couplings and the Yukawa couplings to ensure that they remain
perturbative at electroweak energy scales and all the way up to the unification
scale ($10^{16}$ GeV) or the Planck scale ($10^{19}$ GeV).  Thus, the lowest
energy scales at which the Landau poles are permitted to occur can be taken as
these scales to obtain corresponding electroweak-scale upper bounds on the
values of the Yukawa couplings for the exotic fermions. 

The beta functions of the quartic couplings of the type $b_i$ in the scalar
potential enumerated in Sec.~IV receive negative contributions from the exotic
Yukawa couplings. This can lead to the quartic couplings running to negative
values at energy scales much lower than the unification scale, making the
vacuum unstable at these lower scales.  Some new physics beyond the \esix~
framework would be necessary to restore the vacuum stability. In order to avoid
this and preserve the \esix~features up to the unification scale, the quartic
couplings will have lower bounds at the EW scale such that they do not turn
negative at energies below the unification scale.  The RGE of the quartics is
necessary to compute these lower bounds. These quartics also have an upper
bound due to the presence of the $b_i^2$ terms with positive coefficients in
the beta function which can lead to poles below the unification scales.
The details of these calculations for exotic Yukawa and quartic couplings are
given in Appendix C. We present the main results here.

We find that it is hard to push the upper bound of the vector-like quark mass
above 750 GeV, if we demand perturbative Yukawa coupling constants at least
up to $10^{16}$ GeV. The lower bounds on the vector-like quark masses set by
CMS and ATLAS at the LHC are close to, but higher than, 750 GeV and depend on
the branching ratio to $W+SM$ and $H+SM$~\cite{Khachatryan:2015gza,Aad:2015mba}.
Thus, in order for this model to be a viable theory up to the GUT scale, it is
necessary to avoid these bounds.

This is achieved if the vector-like particles are protected by a $\mathbb{Z}_2$
symmetry, which forbids vertices such as $DqW$, $DqZ$ and $DqH$. This can lead
to another problem. In the absence of these decays, the exotic quarks are
subject to the cross section bounds on long-lived charged particles. As
discussed in Sec.~IX.A, these bounds are even higher than those placed by the
vector-like quark searches. This problem is circumvented by $\mathbb{Z}_2$-%
preserving decays as follows.

As explained in Sec.~IX.A, $\mathbb{Z}_2$-preserving decays of the exotic
quarks to the SM quarks and $\tilde{\nu}$ or $\tilde{N}^c$ allow one to escape
the long-lived charged particle bounds.  They open up a window of viable exotic
quark masses with a lower bound of $\sim 400$~GeV as discussed in Sec.~IX.A.
Demanding the unification-scale perturbativity of the Yukawa couplings implies
an upper bound of $1.3$ on the exotic Yukawa couplings at the electroweak
scale. As the exotic fermion masses are given by $y_i\langle\tn\rangle$, this
sets the upper bound of this window proportional to $\langle\tn\rangle$.
This upper bound on mass is constrained by the vacuum stability 
considerations for a given mass of $\tn$ as follows. 

From Eq.~(\ref{eqn:appA}), for small $b_3$, $b_4$ and a single scalar
generation, we have
\begin{align}\label{eqn:b17}
b_{17}&=\frac{m^2_{\tn}}{2\langle\tilde{n}\rangle^2}~.
\end{align}
For a lower bound of $b_{17}$ for $b_{17}$ at the EW scale, we have an upper
bound on the allowed $\langle\tn\rangle$ for a given mass of $\tn$. Its value is
\begin{align}
\langle\tn\rangle&=\frac{m_{\tn}}{\sqrt{2b_{17}}}~.
\end{align}
Thus, the exclusion due to the vacuum stability constraint is characterized by
a straight line of slope $1/\sqrt{2b_{17}}$ passing through the origin on the
$\langle\tn\rangle-m_{\tn}$ plot with the area above the line excluded.  The
slope of the line decreases with increasing EW-scale values of the exotic
Yukawa couplings as a larger and larger region is excluded. From the example in
Appendix C we see that for the exotic Yukawa coupling of $0.95$ for quarks and
$0.95$ for the leptons we get the lower bound of $1.4$ on $b_{17}$, which
leads to an upper bound of $\sim 0.63 \,m_{\tn}$ on the SO(10) scalar singlet
$(\tilde{n})$ VEV. More
details of bounds related to $b_{17}$ and its connection to bounds on Yukawa
couplings and therefore the mass spectrum are discussed in Sec. XI.

In this window of~$400$ GeV to $1.3\langle\tn\rangle$, the mass of the SO(10)
16-plet neutral scalar $\tilde{\nu}$ or $\tilde{N}^c$ is constrained to be
almost degenerate with the vector-like quark mass for masses lower than $500$
GeV. This result is obtained by recasting the LHC searches as discussed in
Sec.~IX. Thus, the mass of the SO(10)-singlet scalar constrains the rest of the
spectrum via vacuum stability considerations. 

Finally, the standard-model Higgs coupling is unaffected by the addition of the
new particles, as the additional fermions do not have tree-level Yukawa
couplings to the doublet Higgs boson. Thus the danger of the doublet Higgs
quartic coupling going to negative values at the TeV scale is avoided. The
mixing of the Higgs doublet with the singlet may lead to positive contributions
to the doublet quartic beta functions. As discussed in Sec.~IV, this mixing is
constrained to be a small value by the experimental bounds and will not pose a
problem for the stability of the doublet Higgs quartic coupling.
\bigskip

\centerline{\bf VI.  DECAYS OF $\tn$ TO $\gamma\gamma,~\gamma Z,~Z Z,~W^+ W^-,
~ZH$}
\bigskip

The coupling of $\tn$ to a loop of the exotic lepton $L$ generates decays to
other pairs of electroweak gauge bosons besides $\gamma \gamma$, namely $Z
\gamma$, $ZZ$, and $WW$.  For general discussions of the ratio of the
correpsonding partial widths see \cite{Franceschini:2015kwy,Altmannshofer:%
2015xfo,Craig:2015lra,Low:2015qho,Kamenik:2016tuv}.
The gauge-invariant terms in the Lagrangian describing the most general
couplings of $\tn$ to electroweak bosons $W$ and $B$ may be written
\beq
\delta {\cal L}_{\rm ew} =  \frac{\alpha_{\rm em}}{4 \pi x}\frac{\kappa_W}
{4 m_{\tn}} \tn W^a_{\mu \nu} W^{a~\mu \nu}+\frac{\alpha_{\rm em}}{4 \pi (1-x)}
\frac{\kappa_B}{4 m_{\tn}} \tn B_{\mu \nu} B^{\mu \nu}~, x \equiv \sin^2
\theta_W.
\eeq
With $Z = W^3 \cos \theta_W - B \sin \theta_W$, $A = W^3 \sin \theta_W + B
\cos \theta_W$, the couplings are
\beq
g_{\tn \gamma \gamma} = C_0 (\kappa_W + \kappa_B)~,
\eeq
\beq
g_{\tn Z \gamma} = C_0 \sqrt{x(1-x)}\left( \frac{\kappa_W}{x} - \frac{\kappa_B}
{1-x} \right)~,
\eeq
\beq
g_{\tn ZZ} = C_0 \left( \kappa_W \frac{1-x}{x}+\kappa_B \frac{x}{1-x} \right)~,
\eeq
\beq
g_{\tn WW} = C_0 \left( \frac{\kappa_W}{x} \right)~,
\eeq
where $C_0$ is a common factor.  Taking account of the contribution of three
charged vector-like $L$ states in the loop, the ratio of $Z \gamma$ and $\gamma
\gamma$ couplings is found (cf.\ Table \ref{tab:va0}) to be
\beq
\frac{g_{\tn Z \gamma}}{g_{\tn \gamma \gamma}} = [x(1-x)]^{-1/2}\left(
\frac12-x \right)~,
\eeq
consistent with $\kappa_B = \kappa_W$.  Adding the contribution of three $Q =-1/3$ weak isoscalar quarks (``$D_{1,2,3}$''), one finds instead $\kappa_B =
(5/3) \kappa_W$.  In this case, substituting $x = 0.2315$, squaring
amplitudes, and multiplying by 2 for non-identical particles in the final
state, one finds partial decay rates to be in the ratio
\beq \label{eqn:ratios}
\Gamma_{\gamma \gamma}:~\Gamma_{Z \gamma}:~\Gamma_{ZZ}:~\Gamma_{WW} =
1:~0.24:~2.08:~5.32~.
\eeq

% This is Figure 2
\begin{figure}
\begin{center}
\includegraphics[width=0.77\textwidth]{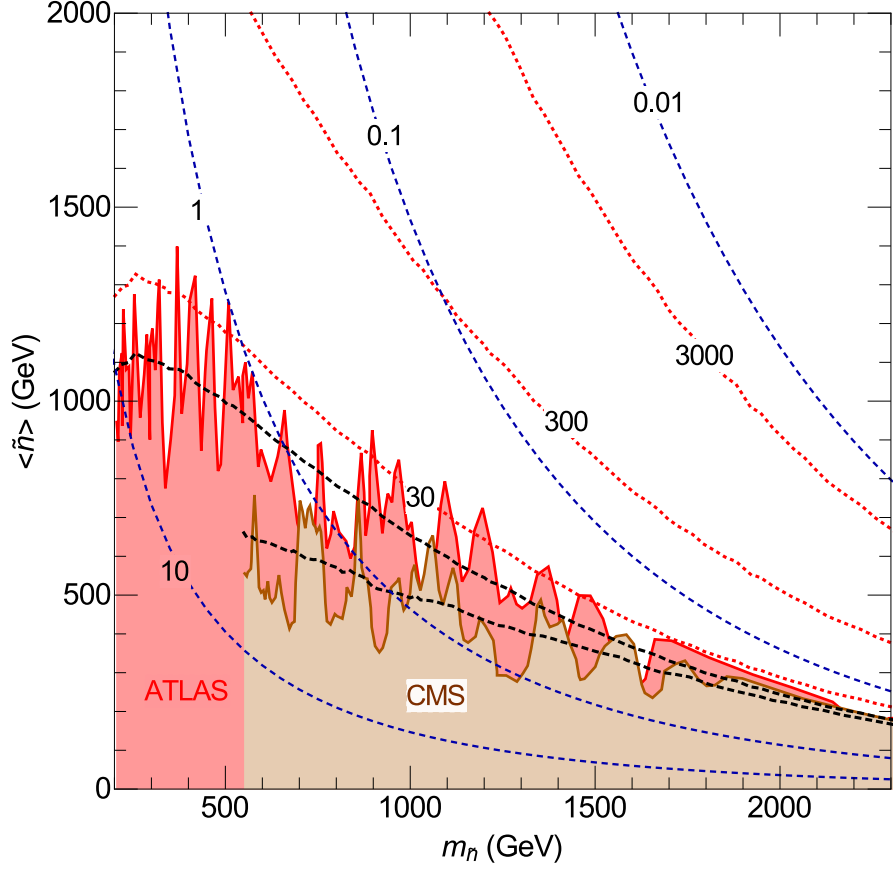}
\end{center}
\caption{Dark [red] and lighter [dark orange] shaded regions show the exclusion
regions for exotics by ATLAS \cite{ATLAS:2016eeo,ATNOGG} and CMS
\cite{Khachatryan:2016yec}, respectively, in the
diphoton channel. \textbf{Black (dashed):} Expected $95\%$ exclusion limit for
ATLAS (upper curve, 15.6 fb$^{-1}$ at 13 TeV) and CMS (lower curve, 16.2
fb$^{-1}$ at 13 TeV $+$ 19.7 fb$^{-1}$ at
8 TeV). \textbf{Red (dotted)}: Contours showing ATLAS reach for non-discovery
for luminosities of 30, 300 and 3000 fb$^{-1}$ at 13 TeV. \textbf{Blue
(dashed):} Contours of constant cross section $\sigma_0 \equiv \sigma(pp \to
\tn X \to \gamma \gamma X)$ in the plane of $\langle \tn \rangle$ versus
$m_{\tn}$ for values of $\sigma_0$, top to bottom, ranging from 0.01 to 10 fb.
\label{fig:cont13}}
\end{figure}
\bigskip

Similar results (aside from a factor of 2 lower for $Z \gamma$) were obtained
in Ref.\ \cite{Csaki:2016kqr}. These ratios should be targets for discovery or
bounds when the LHC accumulates more data at 13 TeV. As shown in Fig.~\ref%
{fig:cont13} and explained in Sec.~VII, $\sigma(pp\rightarrow\tn\rightarrow
\gamma\gamma)$ is excluded for values greater than $1$ fb due to ATLAS (15.6
fb$^{-1}$ at 13 TeV) \cite{ATLAS:2016eeo} and CMS (16.2 fb$^{-1}$ at 13 TeV
$+$ 19.7 fb$^{-1}$ at 8 TeV)\cite{Khachatryan:2016yec} searches and vacuum
stability
constraints.  In the present context of probing for the EW and TeV-scale
signatures of the \esix~model at the LHC, we see that the present upper bound
on $\sigma(pp \to \tn X) {\cal B}(\tn \to W^+W^-)$ \cite{ATLASWW} is greater
than 10 fb, corresponding to a $WW/\gamma \gamma$ ratio of at least 10. For
$\sigma(pp \to \tn X \to Z \gamma X)$, an upper limit for a narrow $\tn$ is at
least about 20 fb \cite{CMSZg16}, a factor of $\sim 100$ above expectation
based on Eq.\ (\ref{eqn:ratios}) and the fact that the $\gamma\gamma$
cross section is less than $1$ fb. Therefore, these channels are much less
sensitive to the discovery of $\tn$ if it exists at TeV or sub-TeV scales,
thus making $\gamma\gamma$ the most significant channel for discovery.

An interesting point has been raised in Ref.\ \cite{Bauer:2016ydr}.  It was
found there that a spinless particle $S$ which is a SM singlet can decay to
$Z$ + (Higgs boson) only if it has CP-odd interactions.  If so, the $S \to ZH$
rate could even surpass that for $S \to \gamma\gamma$.
\bigskip

\centerline{\bf VII.  PRODUCTION OF $\tn$ AND $\gamma\gamma$ DECAY}
\bigskip

\begin{figure}
\begin{center}
\includegraphics[width=0.7\textwidth]{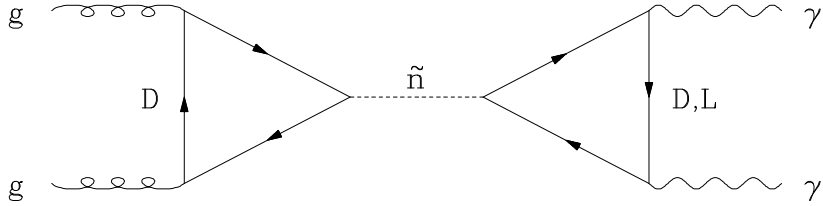}
\end{center}
\caption{Mechanism for hadronic production and decay of scalar boson $\tn$.
\label{fig:proddec}}
\end{figure}

We discuss the production of the scalar member $\tn$ of an \esix~27-plet,
transforming as an SU(5) and SO(10) singlet. Among its decay channels, we focus
on the $\gamma\gamma$ mode as it is the most sensitive to the discovery
of such a singlet among all the decay modes as described in the previous
section.  If $Q_N$ is
respected in its couplings, it does not couple to pairs of ordinary fermions in
the 16-plet of SO(10), but only to the exotic fermions $D$ and $L$.  (In
supersymmetric \esix~versions, the $L$ states may be identified as Higgsinos,
while the $\tL$ are the two Higgs doublets \cite{King:2016wep}.) Then $\tn$ can
be produced via gluon-gluon fusion via loops containing $D$ quarks, one for each
of the three families.  It can decay to two photons via $D$ quarks and exotic
leptons $L^\pm$ in loops, as shown in Fig.~\ref{fig:proddec}.  We then need
expressions for $\Gamma(\tn \to gg)$, $\Gamma(\tn \to \gamma \gamma)$, and
the effective two-gluon luminosity in proton-proton collisions.  We shall rely
on the treatment in Ref.\ \cite{Godunov:2016kqn}.  (For a discussion of an
alternative mechanism for $\tn$ production via the $\gamma \gamma$ initial
state see, e.g., Refs.\ \cite{Csaki:2015vek,Fichet:2015vvy,Fichet:2016pvq,%
Csaki:2016raa,Ghosh:2016lnu,Agarwalla:2016rmw}.)

% This is Figure 3

\begin{figure}
\begin{center}
\includegraphics[width=0.325\textwidth]{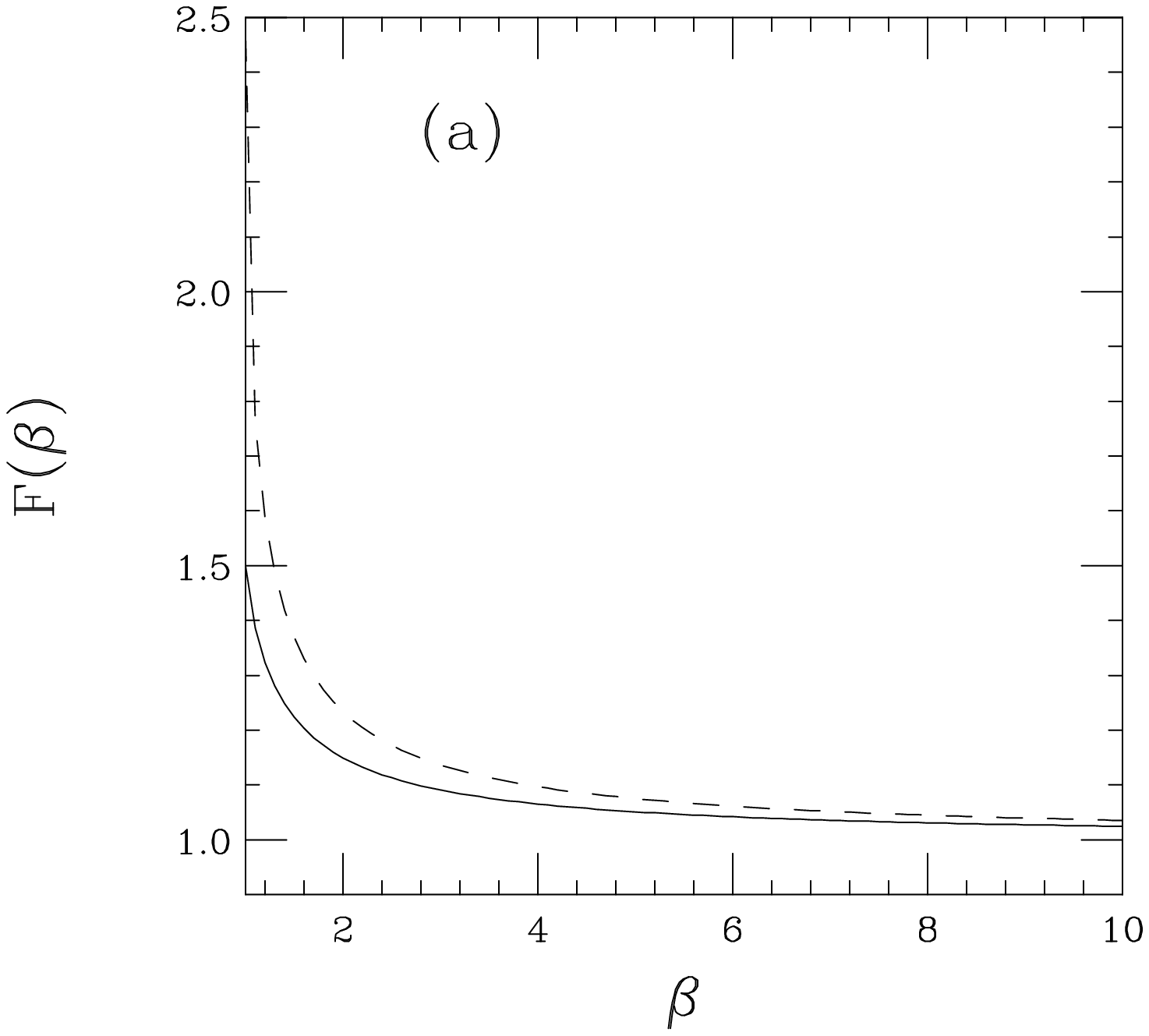}\,\,
\includegraphics[width=0.325\textwidth]{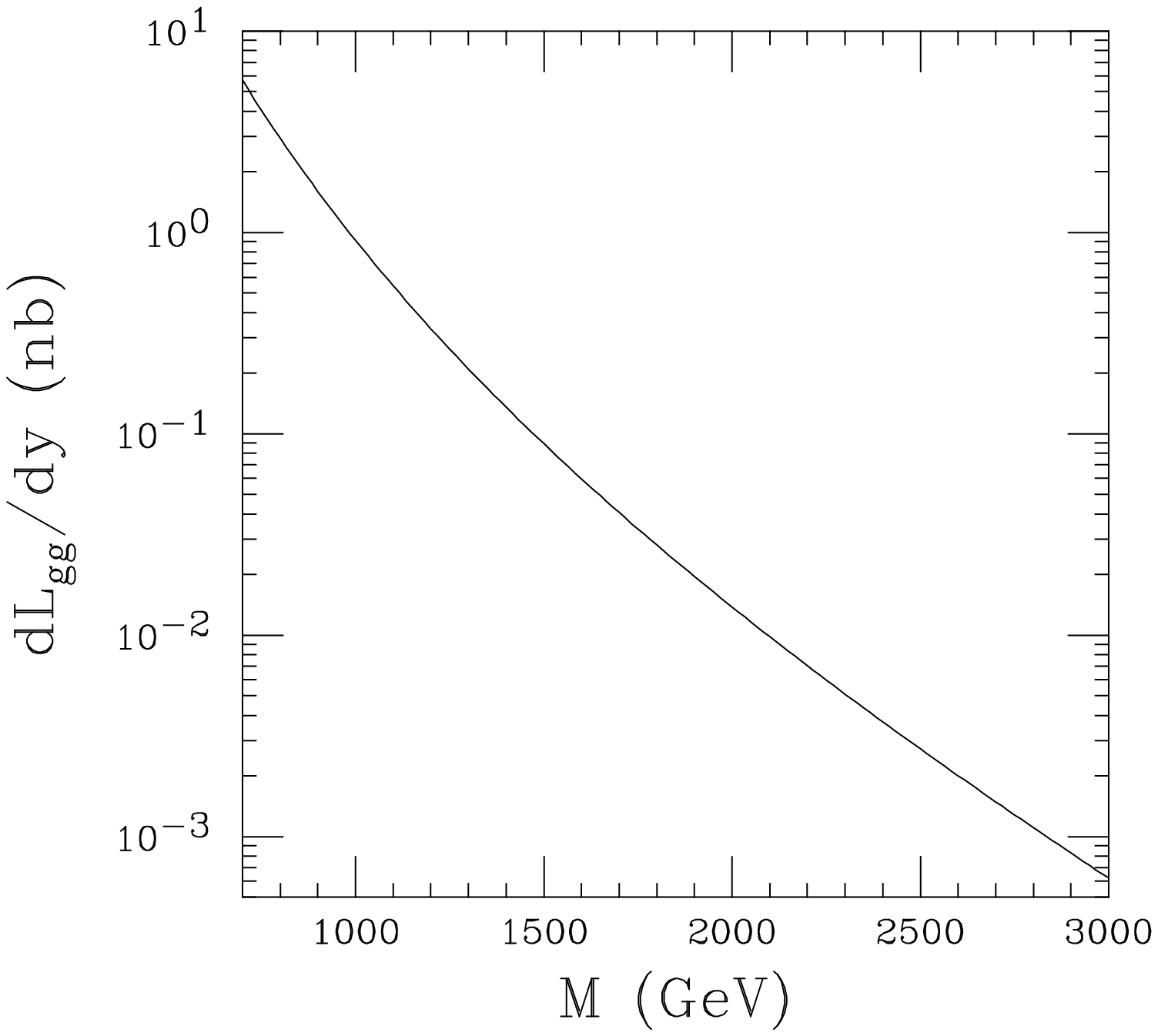}\,\,
\includegraphics[width=0.31\textwidth]{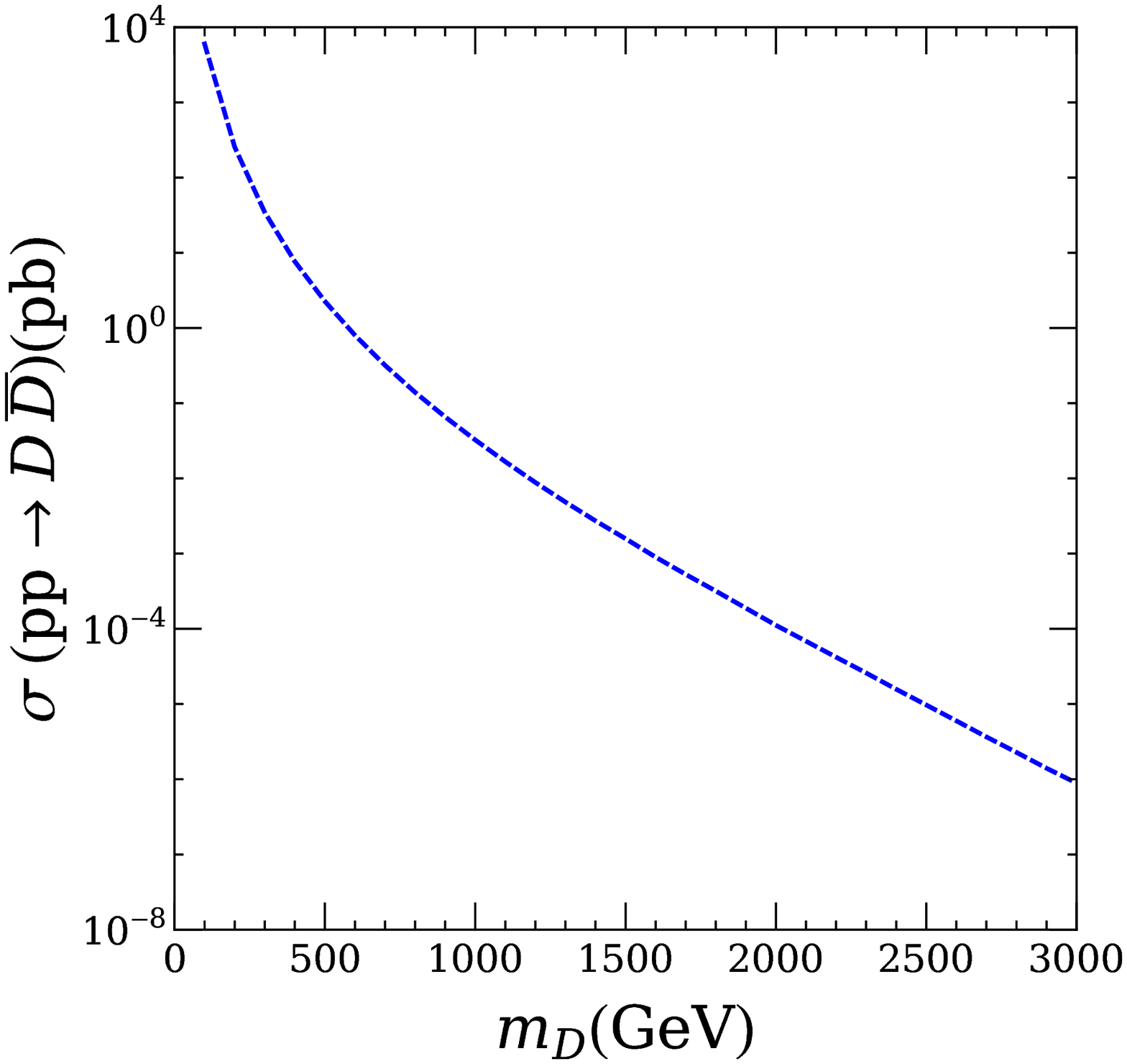}
\end{center}
\caption{Left: Functions $F_\pm(\beta)$ governing decays of scalar
($F_+$) or pseudoscalar ($F_-$) to two photons.  Solid: scalar $\tn$; dashed:
pseudoscalar $\tn$. Middle: Effective gluon-gluon luminosity function
for proton-proton production of a state with mass $M$ at center-of-mass energy
$\sqrt{s} = 13$ TeV. Right: Cross section for $p p \to D D^c + X$ via
two-gluon intermediate state at $\sqrt{s} = 13$ TeV.}
\label{fig:F}
\end{figure}

Two of the neutral companions of the scalar $\tn$
in the 27-plet are the Higgs bosons in a conventional two-Higgs-doublet model.
We assume each $D_i$, where $i=1,2,3$ is the family label, is coupled to
$\tn$ via a term $\Delta {\cal L} = y_{D_i} \tn \bar D_i D_i$.  One then finds
\beq
\Gamma(\tn \to gg) = \left( \frac{\alpha_s}{6 \pi} \right)^2 \frac{m_{\tn}^3}
{2 \pi} \left( \sum_{i=1}^3 \frac{y_{D_i}}{m_{D_i}} F_+(\beta_i) \right)^2~,
\eeq
where $\beta_i \equiv (2 m_{D_i}/m_{\tn})^2$, and for a scalar $\tn$
\cite{Resnick:1973vg,Ellis:1975ap},
\beq \label{eqn:fp}
F_+(\beta) \equiv \frac{3}{2}\beta \left[ 1 + (1-\beta) \arcsin^2
\frac{1}{\sqrt{\beta}} \right]~.
\eeq
For a pseudoscalar $\tn$ (see, e.g., \cite{Djouadi:2005gi,Low:2015qep}):
\beq \label{eqn:fm}
F_-(\beta) = \beta \arcsin^2 (1/\sqrt{\beta})~.
\eeq
These functions are plotted in the left panel of Fig.~\ref{fig:F}. Both $F_+
(\beta)$ and $F_-(\beta)$ approach 1 for large $\beta$.  We do not consider the
case $\beta <1$, in which $\tn \to D \bar D$ becomes kinematically allowed.

We now assume that each $D_i$ obtains its mass through the vacuum expectation
value (VEV) $\langle \tn \rangle$ of $\tn$ itself.  This is the only neutral
27-plet member whose VEV can give mass to the exotic fermions $D$ and $L$, as
noted at the end of Sec.\ II.  In this case one has $y_{D_i}/m_{D_i} = 1/
\langle \tn \rangle$, and the expression for $\Gamma(\tn \to gg)$ reduces to
\beq \label{eqn:G2g}
\Gamma(\tn \to gg) = \left( \frac{\alpha_s}{6 \pi} \right)^2 \frac{m_{\tn}^3}
{2 \pi \langle \tn \rangle^2} \left( \sum_{i=1}^3 F_\pm(\beta_i) \right)^2~.
\eeq
While QCD corrections to this partial width are appreciable --- about a factor
of two \cite{Godunov:2016kqn} --- they will largely cancel out when we express
the cross section for $pp \to \tn X \to \gamma \gamma X$ in terms of the
partial width $\Gamma(\tn \to \gamma \gamma)$.

% This is Figure 4

We next evaluate the cross section for $\tn$ production in $pp$ collisions at
the LHC.  Ref.\ \cite{Godunov:2016kqn} defines a gluon-gluon luminosity as
an integral over the rapidity $y$ at which $\tn$ is produced:
\beq \label{eqn:lgg}
\frac{dL_{gg}}{d \hat s}|_{\hat s = m_{\tn}^2} \equiv \frac{1}{s}
\int_{\ln \sqrt{\tau}}^{\ln 1/\sqrt{\tau}} dy~g(x_1,Q^2)~g(x_2,Q^2)~,
\eeq
where $x_1 \equiv \sqrt{\tau}e^y$, $x_2 = \tau/x_1$, $\tau \equiv m_{\tn}^2/s$,
and $s$ is the square of the total CM energy.  Here $g(x,Q^2)$ is the gluon
structure function, which we take from the CTEQ14 NNLO set \cite{Dulat:2015mca}.
The middle panel of Fig.~\ref{fig:F} shows $dL_{gg}/dM^2 = dL_{gg}/d \hat{s}$ for production of
a state with with mass $M = \sqrt{\hat s}$ in proton-proton collisions at
$\sqrt{s} = 13$ TeV.  This function behaves roughly as $M^{-6}$ in the region
of interest.

The relation between the gluon-gluon luminosity and the cross section
for $\tn$ production in proton-proton collisions may be written as
\beq
\sigma(pp \to \tn X) = \frac{\pi^2}{8 m_{\tn}}\Gamma(\tn \to gg)
\frac{dL_{gg}}{d \hat s}|_{\hat s = m_{\tn}^2}~.
\eeq
Assuming that the total width of $\tn$ is dominated by its two-gluon decay,
this may then be used to calculate the cross section for $\tn$ production and
decay to two photons:
\beq \label{eqn:2gprod}
\sigma(pp \to \tn X \to \gamma \gamma X) = \frac{\pi^2}{8 m_{\tn}}
\Gamma(\tn \to \gamma \gamma) \frac{dL_{gg}}{d \hat s}|_{\hat s = m_{\tn}^2}~.
\eeq

We shall assume that the charged leptons $L_i$ in the right-hand loop of Fig.\
\ref{fig:proddec} have the same masses as the $D_i$, and that there are
three families of them.  We shall also assume that these masses are high
enough above $m_{\tn}/2$ that the functions $F_\pm(\beta)$ may be approximated
by 1.  Then, adapting Eq.\ (9) of Ref.\ \cite{Godunov:2016kqn} to our
assumptions, we find
\beq \label{eqn:G2gam}
\Gamma(\tn \to \gamma \gamma) = \left( \frac{\alpha}{3 \pi} \right)^2
\frac{m_{\tn}^3}{16 \pi \langle \tn \rangle^2} \left[ 3 \cdot 3 \cdot
\left( -\frac{1}{3} \right)^2 + 3 \cdot \left( -1 \right)^2 \right]^2~.
\eeq
The first term in the square brackets is the contribution of the three $D$
quarks, while the second term is the contribution of the three charged leptons
$L$. There is an additional factor of three in the first term as the quarks are
colored.  The branching fraction ${\cal B}(\tn \to \gamma \gamma)$ is
then approximately
\beq \label{eqn:bfgg}
\frac{\Gamma(\tn \to \gamma \gamma)}{\Gamma(\tn \to gg)} = \left(
\frac{\alpha}{\alpha_s} \right)^2 \frac{8}{9} = 5.9 \times 10^{-3}~.
\eeq
(This value will be further reduced, possibly by as much as a factor of two,
by QCD corrections to the denominator.)  Csaki and Randall obtain a similar
value as they have similar contributions to the loop diagrams governing
$\tn \to \gamma \gamma$ \cite{Csaki:2016kqr}.
 
Combining Eqs.\ (\ref{eqn:2gprod}) and (\ref{eqn:G2gam}), we find
\beq \label{eqn:comb2gp}
\sigma(pp \to \tn X \to \gamma \gamma X) = \frac{\alpha^2}{72\pi}
\frac{m_{\tn}^2}{\langle \tn \rangle^2}\frac{dL_{gg}}{d \hat{s}}|_{\hat s =
m_{\tn}^2}~.
\eeq
Contours of equal $\sigma_0 \equiv \sigma(pp \to \tn X \to \gamma \gamma X)$
in the plane of $\langle \tn \rangle$ versus $m_{\tn}$ are easily plotted by
solving Eq.\ (\ref{eqn:comb2gp}) for $\langle \tn \rangle$:
\beq
\langle \tn \rangle = \alpha m_{\tn} \left( \frac{dL_{gg}/d\hat{s}}{72 \pi
\sigma_0} \right)^{1/2}\label{eqn:vacsigma}
\eeq
and varying $m_{\tn}$.  The results for values of $\sigma_0$ between 0.01 and
10 fb are shown in Fig.\ \ref{fig:cont13}, along with experimental limits from
ATLAS \cite{ATLAS:2016eeo} and CMS \cite{Khachatryan:2016yec}.  The vacuum
expectation value $\langle \tn \rangle$ needed to produce a given cross section
$\sigma_0$ varies roughly as $m_{\tn}^{-2}$ for the range shown.
\bigskip

\centerline{\bf VIII.  THE ROLE OF A \z2~SYMMETRY IN NEUTRINO MIXING}
\bigskip

The fundamental 27-plet of \esix~contains five neutral members, whose
left-handed states we have denoted as $[\nu, N^c, L_1^0, L_2^{0c}, n]$
(cf.\ Table \ref{tab:listf}, but omitting the family index $i$).  In
Ref.\ \cite{Rosner:2014cha} we discussed a general $5 \times 5$ mass matrix
in this basis space:

\beq \label{eqn:r5}
{\cal M}_5 = \left[ \begin{array}{c c c c c} 0 & m_{12} & 0 & M_{14} & 0 \cr
 m_{12} & M_{22} & 0 & m_{24} & 0 \cr 0 & 0 & 0 & M_{34} & m_{35} \cr
 M_{14} & m_{24}& M_{34} & 0 & m_{45} \cr 0 & 0 & m_{35} & m_{45} & 0
\end{array} \right]~,
\eeq
where masses with small letters correspond to $\Delta I_L = 1/2$ while
those with capital letters correspond to $\Delta I_L = 0$.  After
diagonalization with respect to the third and fourth rows and columns, this
becomes
\beq \label{eqn:r5p}
{\cal M}'_5 = \left[ \begin{array}{c c c c c}
 0 & m_{12} & M_{14}/\s & M_{14}/\s & 0 \cr
 m_{12} & M_{22} & m_{24}/\s & m_{24}/\s & 0 \cr
 M_{14}/\s & m_{24}/\s & M_{34} & 0 & (m_{35}+m_{45})/\s \cr
 M_{14}/\s & m_{24}/\s & 0 & -M_{34} & (m_{45}-m_{35})/\s \cr
 0 & 0 & (m_{35}+m_{45})/\s & (m_{45}-m_{35})/\s & 0
\end{array} \right]~.
\eeq
The first two rows and columns correspond to states with \z2~$=-1$, while the
last three correspond to states with \z2~$=+1$.  In the limit of exact
\z2~symmetry, the parameters $M_{14}$ and $m_{24}$ vanish, so ${\cal M}'_5$
reduces to the direct sum of $2 \times 2$ and $3 \times 3$ matrices.  The
$2 \times 2$ matrix corresponds to the standard seesaw picture, with $M_{22}$
taking on a large value to force SM neutrinos to have small masses
$-m_{12}^2/M_{22}$.  The $3 \times 3$ matrix has two large eigenvalues
$\pm M_{34}$ (pseudo-Dirac $L^0_{1,2}$ states) and one small eigenvalue which
we may identify as the $n$ mass:
\beq
m_n = - 2 m_{35} m_{45} / M_{34}~.
\eeq

Here $M_{34}$ may be as light as several hundred GeV, while $m_{34}$ and
$m_{45}$ should be of the same order as SM quark and lepton masses.  Thus
the $n$ states of each family should be lighter than the SM fermions in
that family, but could be considrably heavier than the corresponding neutrinos.

In Ref.\ \cite{Rosner:2014cha} the states $n$ were proposed as sterile neutrino
candidates coupling to SM neutrinos through a violation of the \z2~symmetry, in
order to explain various apparent anomalies in the three-active-neutrino
picture. In the scenario in which \z2~is exact, however, a $n$ state cannot
account for the above-mentioned anomalies.

An exact \z2~symmetry could account for the existence of dark matter, in
the form of the lightest state with \z2 = --1.  The scalar $\tnu$ or $\tN^c$
could be one such candidate.
Other heavier states with \z2 = --1 could decay to it and one or several states
with \z2 = 1. A full discussion of these possibilities is beyond the scope
of the present paper, but some examples will be given in the next Section.
\bigskip

\centerline{\bf IX.  SIGNATURES FOR SO(10) 10-PLET FERMIONS}
\bigskip

The decays of exotic quarks and leptons in the 10-dimensional representation of
SO(10) depend crucially on whether the \z2~symmetry defined earlier is
approximate or exact.  We recall that singlets and 10-plets of SO(10) are
assigned \z2~= +1 while 16-plets of SO(10) are assigned \z2~= --1. If $\langle
\tnu \rangle = \langle \tN \rangle = 0$, the \z2~is exact, while if one or both
of these VEVs is non-vanishing, exotic 10-plets can mix with SM 16-plets.
\bigskip

\leftline{\bf A.  Exotic quark production}
\bigskip

Estimates of $D$ pair production were made in Refs.\ \cite{Andre:2003wc} and
\cite{Bjorken:2002vt}, among many other places.  We update those predictions
for proton-proton collisions at the LHC center-of-mass (CM) energy of 13
TeV in the right panel of Fig.\ \ref{fig:F}.  If the scalar meson $\tnu$ or
$\tilde{N}^c$ is light enough, its exchange in the reaction $d d^c \to D D^c$
can provide an additional significant contribution to $D$ pair production. 

% This is Figure 5
\begin{figure}
\begin{center}
\includegraphics[width=0.8\textwidth]{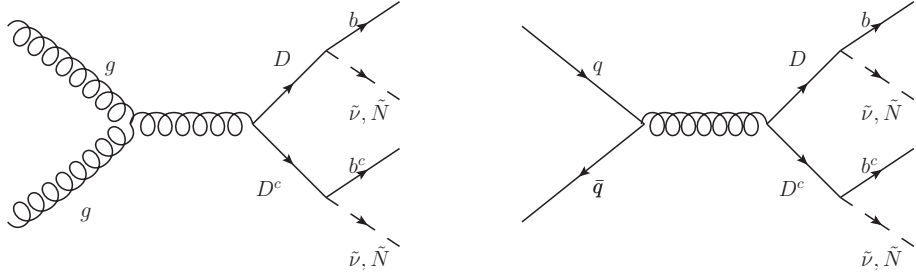}
\end{center}
\caption{Main diagrams contributing to \z2-preserving decays of exotic quark
$D$.}
\label{fig:Dsearch}
\end{figure}

The only couplings that allow for the decay of an exotic $D$ quark in the
\z2-symmetry conserving way are the $\tilde{\nu}Dq_{d,s,b}$ and $\tilde{N}^cD
q_{d,s,b}$ couplings. The inclusive searches for $2b$ jets plus missing
transverse energy (MET) provide experimental lower bounds on the masses of
these exotic quarks. Fig.~\ref{fig:Dsearch} shows the dominant diagrams for the
production and decay of these exotic quarks that contribute to their
experimental searches. 
 
For these decays to be kinematically viable, we need $m_{\tilde\nu}$ and/or
$m_{\tilde{N}^c}$ to be less than $m_D$. The ATLAS searches at 8
TeV~\cite{Aad:2015pfx} and 13 TeV~\cite{Aaboud:2016nwl,ATVT,ATSQ}; CMS
searches~\cite{Khachatryan:2015wza} in Run-1 data for third generation squarks;
and more recent CMS searches \cite{CMS17001,CMS16008,CMS16032,CMS16051}
put stringent lower bounds on the masses. Near the
limit in which the difference between $m_{\tilde{\nu},\tilde{N}^c}$ and $m_D$
is close to zero, the $2b$ jets plus MET searches are not sensitive.  However,
for exotic quark masses below $400$ GeV, the monojet searches are sensitive in
this limit, excluding exotic quarks in this region of parameter space.

For $m_D>400$ GeV, a considerable region of the parameter space is allowed and
grows with the mass of the exotic quark. In order to estimate this difference,
the $95\%$ $CL$ exclusion bound in~\cite{Aaboud:2016nwl}, which is the most
stringent bound on the sbottom searches, is recasted. As the $gg\rightarrow
D\overline{D}$ and $q\overline{q}\rightarrow D\overline{D}$ cross section is
higher than the $gg\rightarrow \tilde{b}\tilde{\overline{b}}$ and
$q\overline{q}\rightarrow b\tilde{\overline{b}}$ cross sections and as there
are three copies of the exotic
quarks, the production cross section of these quarks is greater than that of
the sbottoms by about a factor of 20 near the electroweak and TeV scales.  This
results in more parameter space being excluded at $95\%$ $CL$ compared to the
sbottom searches. We obtain results very close to such a recast performed
in~\cite{Kawamura:2016idj}, which leads to an allowed mass difference of about
$200$ GeV between $m_D$ and $m_{\tilde{\nu},\tilde{N}^c}$ at $m_D=1$ TeV. For
$m_D<500$ GeV, a near degeneracy between $m_D$ and $m_{\tilde{\nu},
\tilde{N}^c}$ is required to escape jets plus MET searches.  For $m_D<400$ GeV,
this degeneracy is not sufficient to evade the LHC searches as the monojet
searches exclude the presence of exotic quarks in this limit.

As pointed out in~\cite{Carena:2016bnq}, for small decay widths of the
vector-like quarks, quarkonium will form before the decay of the quark for an
energy scale equal to twice the mass of the vector-like quark. This quarkonium
can then decay to produce a peak in the $\gamma\gamma$ spectrum at the mass
twice that of the vector-like quark. The cross section for this decay,
$\sigma(pp\rightarrow DD^c\rightarrow \gamma\gamma)$, is proportional to
$N^2Q^4$, where $N$ is the number of generations and $Q$ is the electric
charge. As quoted in~\cite{Carena:2016bnq}, this cross section is $10$ fb for
the bound state mass of $800$ GeV with $N=1$ and $Q=5/3$ at $\sqrt{s}=13$ TeV.
In our model $N=3$ and $Q=1/3$, which results in the reduction of the cross
section by a factor of $\sim 70$. Therefore, in our case, this cross section at
the bound state mass of $800$ GeV is $\sim 0.14$ fb and decreases exponentially
with increasing bound state mass. As the lower bound on the exotic quarks in
our model is $\sim 400$ GeV as allowed by the LHC searches described above, the
lower bound on the quarkonium mass is $\sim 800$ GeV. Thus, in our model such
a $\gamma\gamma$ peak due to a quarkonium bound state can only appear at
higher energies for higher luminosities than those corresponding to the data
analyzed up to August 2016.

In the limit of preserved \z2~symmetry, the region of parameter space for which
$m_{\tilde{\nu},\tilde{N}^c} > m_D$ implies that the new exotic quarks are
long-lived stable particles. These will be subject to R-hadron searches,
stopped long-lived particle searches, and searches for disappearing tracks at
both ATLAS~\cite{Aad:2013gva,ATLAS:2014fka,Aaboud:2016uth} and CMS
\cite{Chatrchyan:2013oca,CMS:2015kdx,CMS:2014gxa,Khachatryan:2015jha}. In
particular, the ATLAS search~\cite{ATLAS:2014fka} for full Run-1 data excludes
long-lived sbottoms for masses lower than $845$ GeV. 

Given that the production cross section for the exotic quarks in our model is
about 20 times higher than the sbottom pair production in this region, the
experimental lower bound on the exotic quark mass is well above 1 TeV. As shown
in Fig.~\ref{fig:cont13}, the ATLAS~\cite{ATLAS:2016eeo} and
CMS~\cite{Khachatryan:2016yec} searches in the di-photon channel put bounds on
the vacuum expectation values of the $\tilde{n}$ field.  The ATLAS bound
\cite{ATLAS:2016eeo} with 15.6 fb$^{-1}$ (see \cite{ATNOGG} for an update) is
the strongest for a low di-photon decay width of $4$ MeV, and corresponds to
$\langle \tn \rangle = 700$ GeV for $m_{\tn}=1$ TeV.  Naive future projections
assuming constant acceptance for higher luminosities can potentially raise this
bound on the VEV to $1.4$ TeV for an integrated LHC luminosity of $300$
fb$^{-1}$ and to about $2.5$ TeV for $3000$ fb$^{-1}$, if no di-photon
resonance is discovered below 1 TeV.  Thus, the inequality $m_{\tilde{\nu},
\tilde{N}^c} > m_D$ can be viable for exotic quarks heavier than long-lived
search bounds from the LHC.

Finally, the \z2~symmetry can be broken to allow mixing of the exotic quarks
with the SM quarks.  Let us assume that one such quark, called $D_3$, decays
mainly via mixing with the $b$ quark.  Final-state branching ratios are then
predicted to be $2:1:1$ for $Wt:Zb:Hb$, where $H$ denotes the SM Higgs boson
with mass 125 GeV \cite{Andre:2003wc}, when the branching ratio
for $D_3\rightarrow b+m_{\tilde{\nu},\tilde{N}^c}$ is suppressed, which would
be the case for $m_{\tilde{\nu},\tilde{N}^c} > m_D$. The most promising of
these final states is probably $Zb$, where $Z \to e^+e^-,\mu^+ \mu^-$, or $b
\bar b$ (the last identified through $b$-tagging).

Published LHC lower limits on $m_D$ are of order 750 GeV at $\sqrt{s} = 8$ TeV.
Specifically, ATLAS sets lower limits of 755 GeV in the mode $D_3 \to Zb$
\cite{Aad:2014efa} and 735 GeV in the mode $D_3 \to Hb$ \cite{Aad:2015kqa}.
The lower limits set by CMS \cite{Khachatryan:2015gza} range between 740 and
900 GeV depending on the values of the branching fractions of $D_3$ to
$Wt,~Zb,$ and $Hb$. For this case of branching ratios $2:1:1$, the lower bound
on the $D_3$ mass is $790$ GeV. (CMS also searches for vector-like heavy quarks
with charge 2/3 \cite{Khachatryan:2015oba}.) A mass of 750 GeV corresponds to a
cross section at 13 TeV of about 100 fb.
\bigskip

\leftline{\bf B.  Exotic lepton production}
\bigskip

\begin{figure}
\begin{center}
\includegraphics[width=0.52\textwidth]{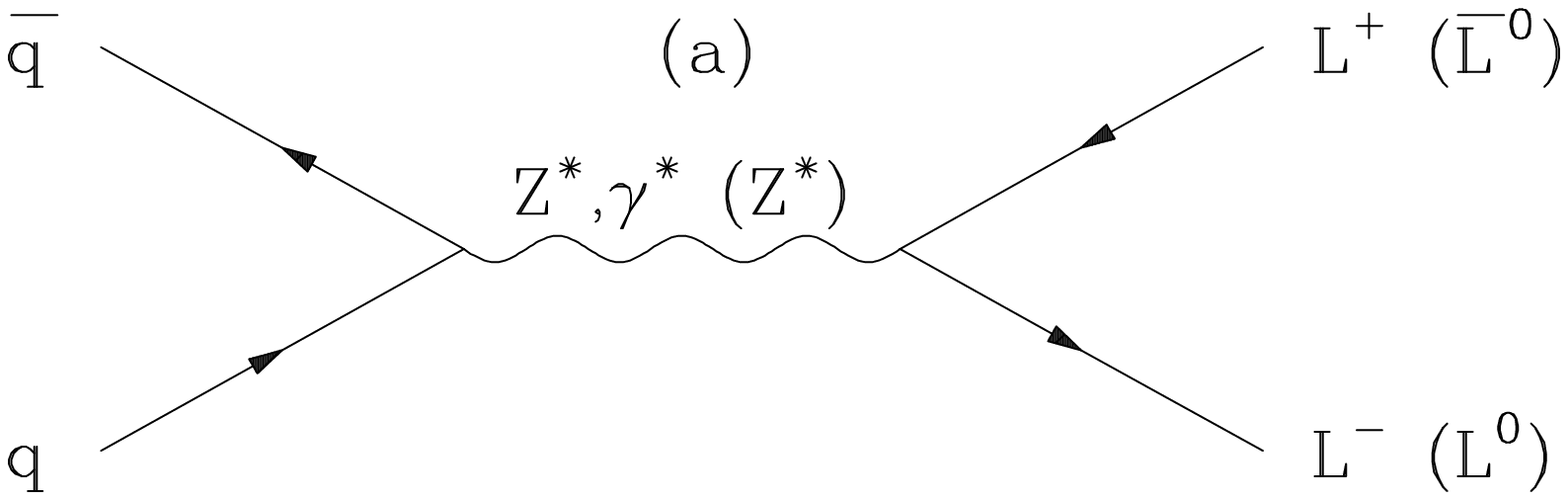}\\
\vskip 0.2in
\includegraphics[width=0.46\textwidth]{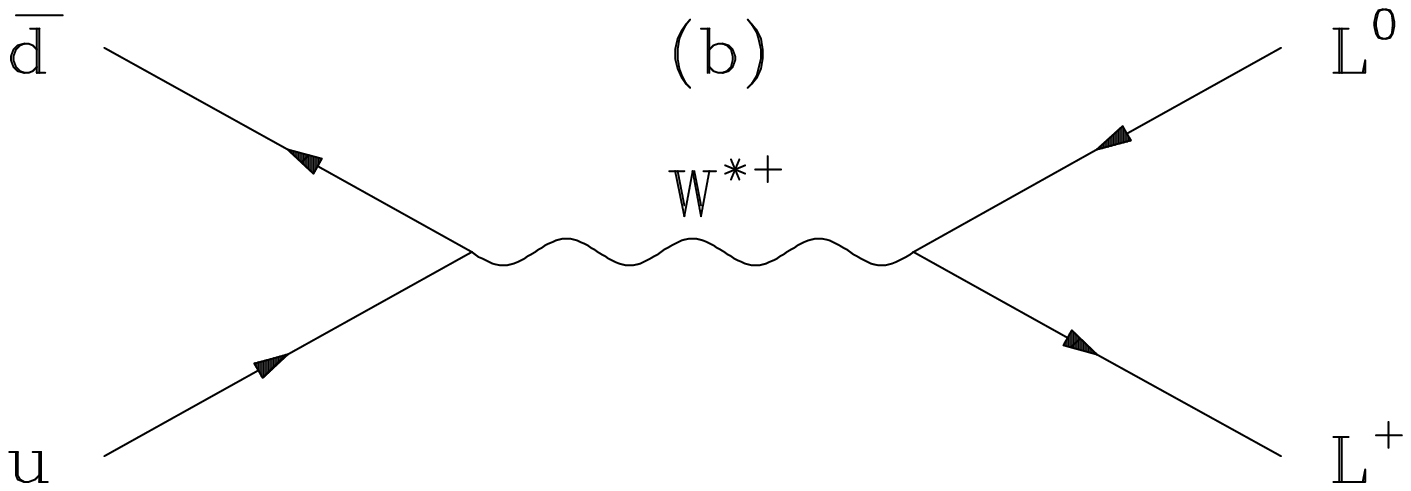}
\hskip 0.2in
\includegraphics[width=0.46\textwidth]{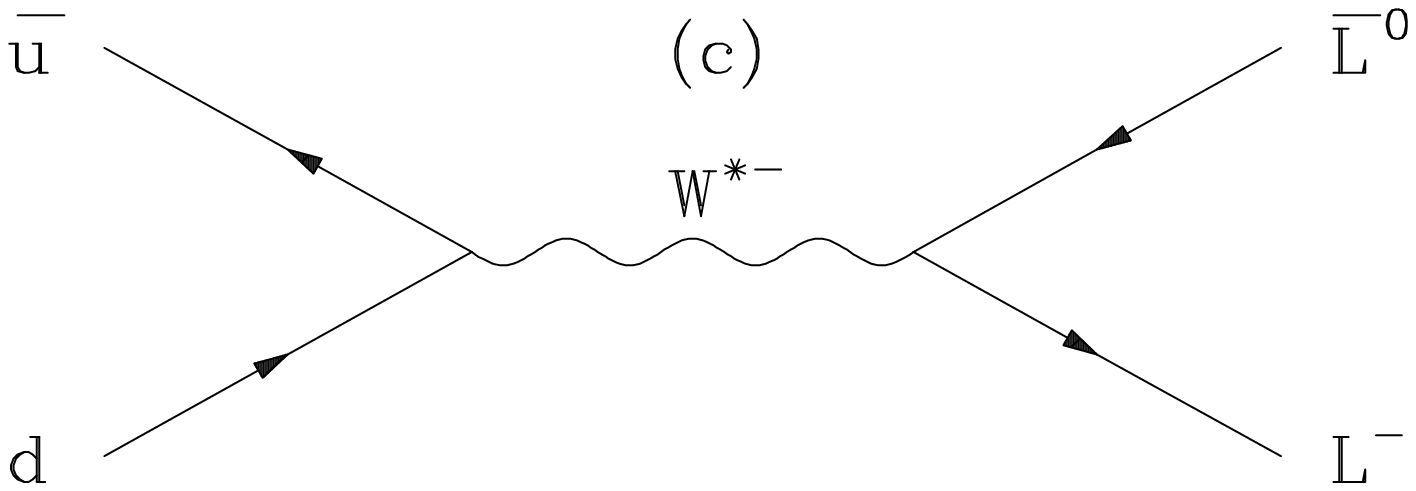}
\end{center}
\caption{Drell-Yan processes contributing to exotic lepton pair production
via (a) virtual $\gamma,~Z^0$; (b) virtual $W^+$; (c) virtual $W^-$.
\label{fig:dy}}
\end{figure}

The weak isodoublet vector-like leptons $L$ can be produced in pairs by the
Drell-Yan process, illustrated in Fig.\ \ref{fig:dy}.  Both a virtual photon
($\gamma^*$) and a virtual SM $Z$ boson ($Z^*$) contribute to $L^+ L^-$
production; only $Z^*$ contributes to $L^0 \bar L^0$ production; and
$W^{* \pm}$ contributes to $L^+ L^0$ or $L^- \bar L^0$ production.

% This is Figure 6

For Drell-Yan production of $L^+ L^-$ and $L^0 \bar L^0$ \cite{Rosner:1986cv},
let $M$ be the effective mass of the $L \bar L$
pair, $y$ its pseudorapidity, and $\theta^*$ the angle between the outgoing
lepton $L$ and the incident quark $q$ in the $q \bar q$ CM.  Let the proton
with laboratory momentum $+p$ ($-p$) emit a parton with momentum fraction
$x_A$ ($x_B$).  Then for vector-like $L \bar L$ production,
\beq \label{eqn:dy}
\frac{d \sigma(pp \to L \bar L + \ldots)}{dM~dy~d(\cos \theta^*)} = \frac{M
x_A x_B}{48 \pi} \left[ \sum_q [f_q^A(x_A) f_{\bar q}^B(x_B)+f_{\bar q}^A(x_A)
f_q^B(x_B) ] S_q (1 + \cos^2 \theta^*) \right]~,
\eeq
where $S_q$, to be defined presently, incorporates the couplings of initial
and final fermions to the virtual photon and $Z$.  In general there would also
be a term proportional to $\cos \theta^*$, but it is absent here because
neither the charged nor the neutral $L$ has an axial-vector coupling to the
$Z^0$.  Specifically, defining
\beq \nonumber
\Delta_\alpha \equiv M^2 - M_\alpha^2~,~~\gamma_\alpha \equiv M_\alpha
\Gamma_\alpha~,~~D_\alpha \equiv (\Delta_\alpha^2 + \gamma_\alpha^2)^{-1}~,
\eeq
\beq
X_{\alpha\beta} \equiv D_\alpha D_\beta(\Delta_\alpha \Delta_\beta
 + \gamma_\alpha \gamma_\beta)~,
\eeq
we have (with Greek letters standing for $\gamma,Z^0$)
\beq
S_q = \sum_{\alpha,\beta} X_{\alpha\beta}(C_V^{q,\alpha} C_V^{q,\beta}
 + C_A^{q,\alpha} C_A^{q,\beta})(C_V^{L,\alpha} C_V^{L,\beta}
 + C_A^{L,\alpha} C_A^{L,\beta})~.
\eeq
The vector and axial-vector couplings of the initial and final fermions are
listed in Table \ref{tab:va0}, where $g_Z \equiv e/\sqrt{x(1-x)},~x \equiv
\sin^2 \theta_W$, and $-e$ is the electron charge.  We consider only
contributions of $u$, $d$, and $s$ partons and antipartons in the proton.
Integrating Eq.\ (\ref{eqn:dy}) over $y$, $\cos \theta^*$, and $M$,
for the example of $M(L) = 400$ GeV, one finds
\beq
\sigma(pp \to L^+ L^- X) = 12.4~(13.6,~16.3)~{\rm fb}~,~~
\sigma(pp \to L^0 L^0 X) = 11.3~(12.7,~15.7)~{\rm fb}~,
\eeq

% This is Table IV
\begin{table}
\caption{Vector and axial-vector couplings of fermions for $L^+ L^-$ and
$L^0 \bar L^0$ production.
\label{tab:va0}}
\begin{center}
\begin{tabular}{c c c c c c c} \hline \hline
 & \multicolumn{6}{c}{Fermion} \\
 & \multicolumn{2}{c}{$u$ quark} & \multicolumn{2}{c}{$d$ quark} 
 & $L^-$ lepton & $L^0$ lepton \\ \hline
Boson    &         $C_V$        &  $C_A$  &        $C_V$       &   $C_A$  &
      $C_V$     &   $C_V$  \\
$\gamma$ &        $2e/3$        &    0    &       $-e/3$       &    0     &
      $-e$      &     0   \\
  $Z^0$  & $g_Z(\frac14-2x/3)$ & $-g_Z/4$ & $g_Z(-\frac14+x/3)$ & $g_Z/4$ &
 $g_Z(-\frac12+x)$ & $g_Z/2$ \\
\hline \hline
\end{tabular}
\end{center}
\end{table}

\noindent
where the first values are based on the above expressions.  The second set are
obtained using Madgraph \cite{Madgraph}, which has subroutines for production
of charginos and neutralinos in the Minimal Supersymmetric Standard Model
(MSSM).  The $L^\pm$ may be identified with Higgs-like charginos, while the
neutral $L$s may be identified with Higgs-like neutralinos.  The third set,
CERN Higgsino cross sections at 13 TeV \cite{CERNLLB}, includes higher-order
corrections (``$K$-factors'') which are seen to be relatively modest.

\begin{figure}
\includegraphics[width=0.45\textwidth]{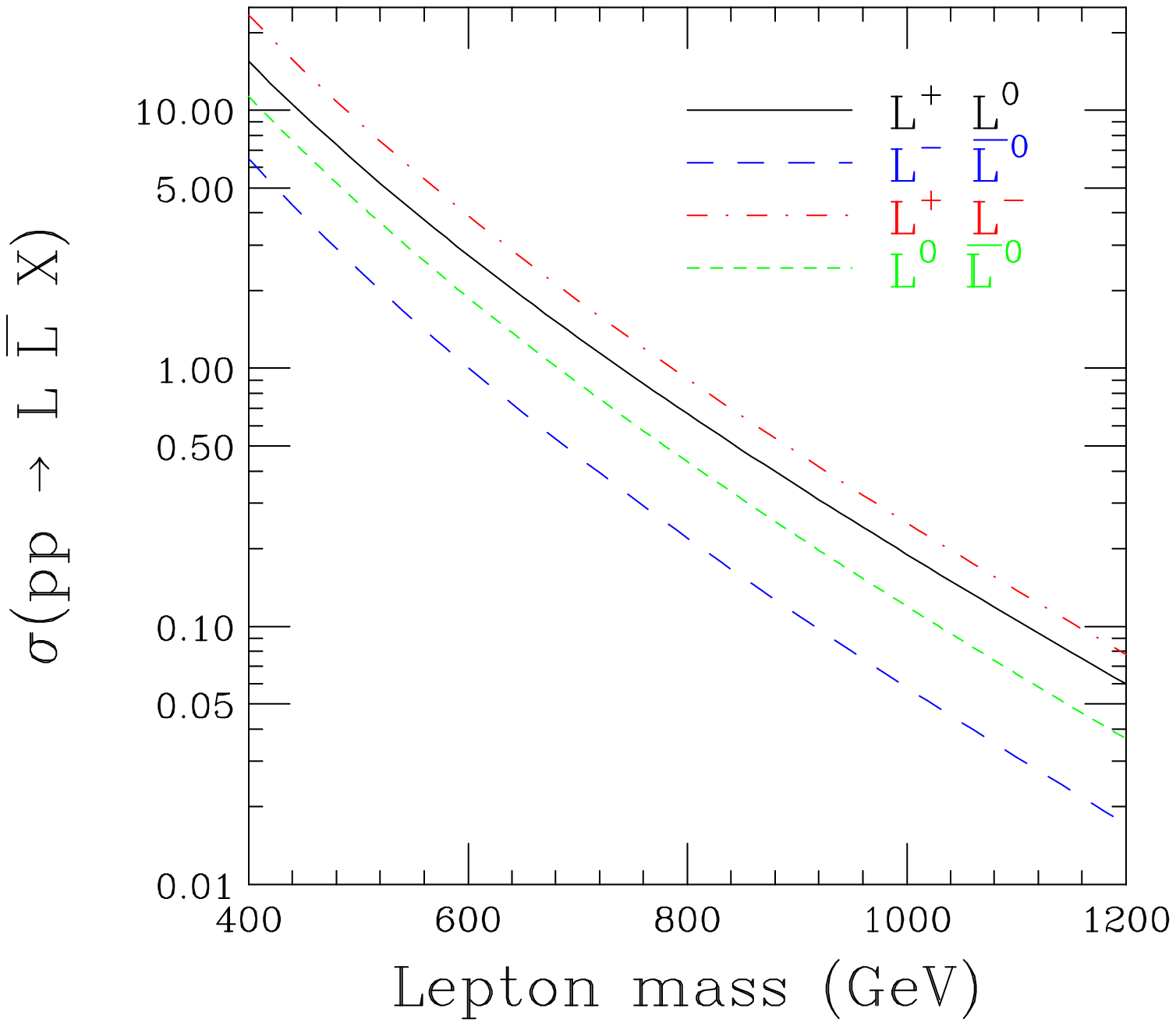}\quad\quad
\includegraphics[width=0.45\textwidth]{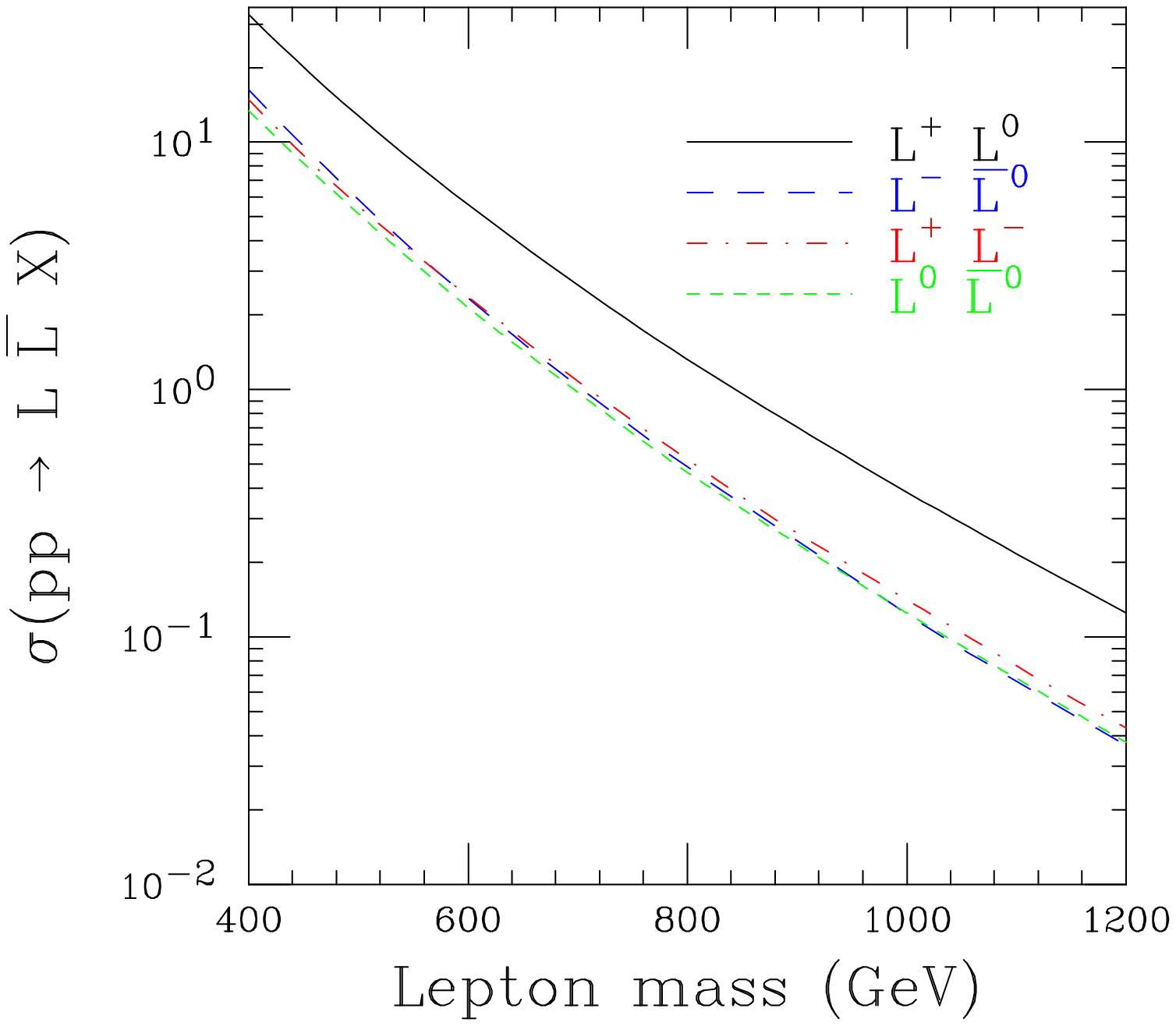}
\caption{Predicted cross sections in fb for production of $L \bar L$ pairs by
proton-proton collisions at $\sqrt{s} = 13$ TeV.  Left:  Leading-order;
right: calculation for Higgsinos using Madgraph \cite{Madgraph}.
\label{fig:dysig}}
\end{figure}

The formalism for production of charged exotic lepton pairs via
virtual $W^\pm$ is similar.  We list the relevant coupling constants in
Table \ref{tab:vapm}.  Here $g_W = e/\sqrt{x}$.  The predicted cross sections
for $L \bar L$ production at 13 TeV are shown in Fig.\ \ref{fig:dysig}.
(We believe the cross sections for charged exotic pairs given in Ref.\
\cite{CERNLLB} are high by a factor of 2.)

Decays of charged and neutral $L$s are problematic. The heavier is likely to
decay via beta-decay to the lighter unless their masses are very close to one
another.  The lighter is likely to decay via mixing with a light lepton
if \z2~symmetry is broken by a small VEV of $\tnu_e$ or $\tn^c$ (see Table
\ref{tab:lists}). Thus, if the neutral $L$ is lighter, we will have such
processes as $L^0 \to \ell_i^- W^+$ and $L^0 \to \nu_i Z^0$, while if
the charged $L$ is lighter, we will have, e.g., $L^- \to \ell_i^- Z^0$
and $L^- \to W^- \nu_i$.  Here $\ell_i^- = (e^-,\mu^-,\tau^-)$ and $\nu_i =
(\nu_e,\nu_\mu,\nu_\tau)$.  If the \z2~symmetry is exact, the charged and
neutral $L$s will decay to a $\tnu$ or a $\tn$ and a SM lepton if kinematically
allowed.

% This is Table V
\begin{table}
\caption{Vector and axial-vector couplings of fermions for $L^+ L^0$ and
$L^- \bar L^0$ Drell-Yan production via $W^{*\pm}$.
\label{tab:vapm}}
\renewcommand{\baselinestretch}{1.4}
\begin{center}
\begin{tabular}{c c c c c c} \hline \hline
\multicolumn{2}{c}{$u\bar d$ } & \multicolumn{2}{c}{$d \bar u$} &
 $L^+ L^0$ & $L^- \bar L^0$ \\ \hline
$C_V$  &  $C_A$  &  $C_V$  &  $C_A$  &  $C_V$  &  $C_V$ \\
$\sqrt{2}g_W/4$ & $-\sqrt{2}g_W/4$ & $\sqrt{2}g_W/4$ & $-\sqrt{2}g_W/4$ &
 $\sqrt{2}g_W/2$ & $\sqrt{2}g_W/2$ \\ \hline \hline
\end{tabular}
\end{center}
\end{table}

% This is Figure 7

The sensitivities of searches for charged and neutral leptons $L$ are highly
dependent on their decay modes, in which mixing with the light charged and
neutral leptons plays a key role.  (See, e.g., \cite{Andre:2003wc}.) Limits are
given by ATLAS \cite{Aad:2013jja,Aad:2015dha,Aad:2015cxa,Grancagnolo:2015,%
Aad:2016giq} and CMS \cite{CMS:2012ra,Chatrchyan:2014aea,CMS:2015mza,%
Khachatryan:2015scf} at $\sqrt{s} = 8$ TeV, and by CMS \cite{CMSL13} at 13
TeV.  The value $M(L^\pm)=400$ GeV which we have quoted above is
near the allowed lower limit \cite{Djouadi:2016eyy}; lower limits on heavy
exotic quark masses are typically twice as large.  The associated production
reactions $p p \to (L^- \bar L^0, L^+ L^0) + X$ via virtual $W$ exchange are a
promising way of producing the exotic heavy leptons \cite{Djouadi:2016eyy}.
Searches should bear in mind the possibility that the exotic leptons of all
three fermion 27-plets may have masses in the several-hundred-GeV range,
giving rise to peaks in $Z \ell$ and $W \ell$ mass distributions.
\bigskip

\leftline{\bf C.  Sterile neutrinos $n_{1,2,3}$}
\bigskip

A recent treatment of sterile neutrinos within the context of \esix~was
given in Ref.\ \cite{Rosner:2014cha}.  At most two such neutrinos are
assigned masses in the eV range to improve fits to oscillations and possible
depletions of reactor fluxes.  This leaves one or two of the $n_i$ to acquire
higher masses, possibly in the keV range as a dark matter candidate to account
for depletion of small-scale structure of the Universe
\cite{Dodelson:1993je,Shi:1998km,Kusenko:2009up,Abazajian:2012ys}
or for a weak gamma-ray line at 3.5 keV stemming from decay of a 7 keV neutrino
\cite{Bulbul:2014sua,Boyarsky:2014jta}.  We shall not discuss the pros and
cons of such an assignment here (see \cite{Adhikari:2016bei} for a thorough
treatment), as we are concerned mainly with hadron collider signatures.

In this context one may note that through imposition of
a discrete \z2~symmetry suppressing the VEVs of the neutral SO(10) 16-plets
$\tnu$ and $\tN^c$, a $n$ mass may be generated entirely through mixing with
the exotic leptons $L^0$ and $\bar L^0$ \cite{Rosner:2014cha}, and so in
principle can be large, even reaching the TeV scale.  Such mixing would affect
the decay schemes of charged and neutral leptons $L$.
\bigskip

\centerline{\bf X.  DIAGNOSTICS FOR EXTRA $Z$ BOSONS}
\bigskip

Many grand unified theories can have neutral gauge bosons heavier than the
$Z^0$ but still accessible at present hadron collider energies \cite{Hewett:%
1988xc,Carena:2004xs,Accomando:2010fz}.  We update a discussion regarding their
identification.  It was shown in Refs.\ \cite{Langacker:1984dc,London:1986dk,%
Rosner:1986cv,Rosner:1989jb,Rosner:1995ft} that a good diagnostic tool
for determining the nature of any $Z$ is the forward-backward asymmetry of
the lepton pairs to which it decays.  In a proton-antiproton collider a
nonzero asymmetry can occur for lepton pairs integrated over the rapidity $y$
of their CM.  For a proton-antiproton collider, this asymmetry is an odd
function of rapidity, so it must be displayed as a function of $y$.

\begin{figure}
\begin{center}
\includegraphics[width=0.48\textwidth]{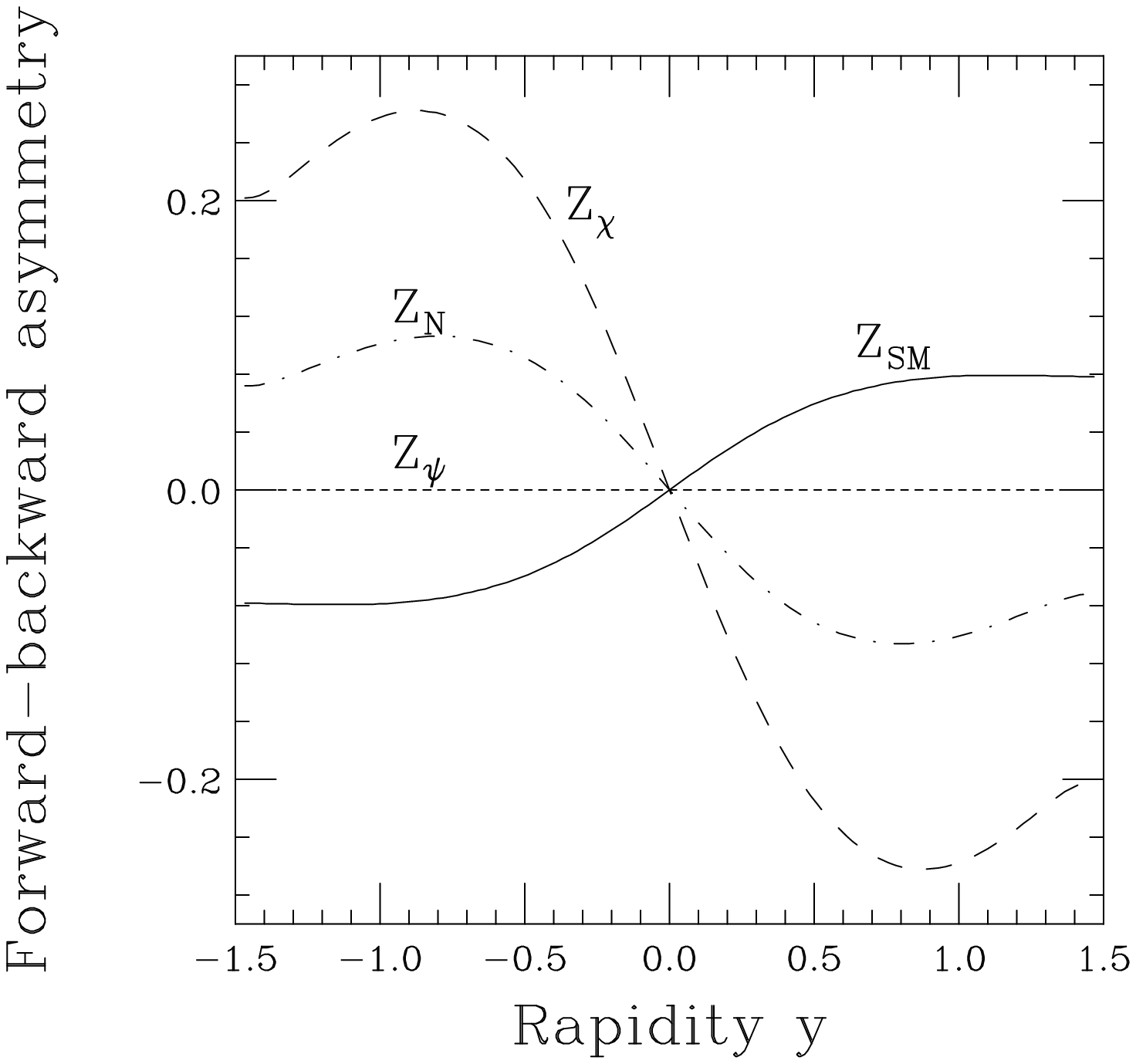}
\includegraphics[width=0.48\textwidth]{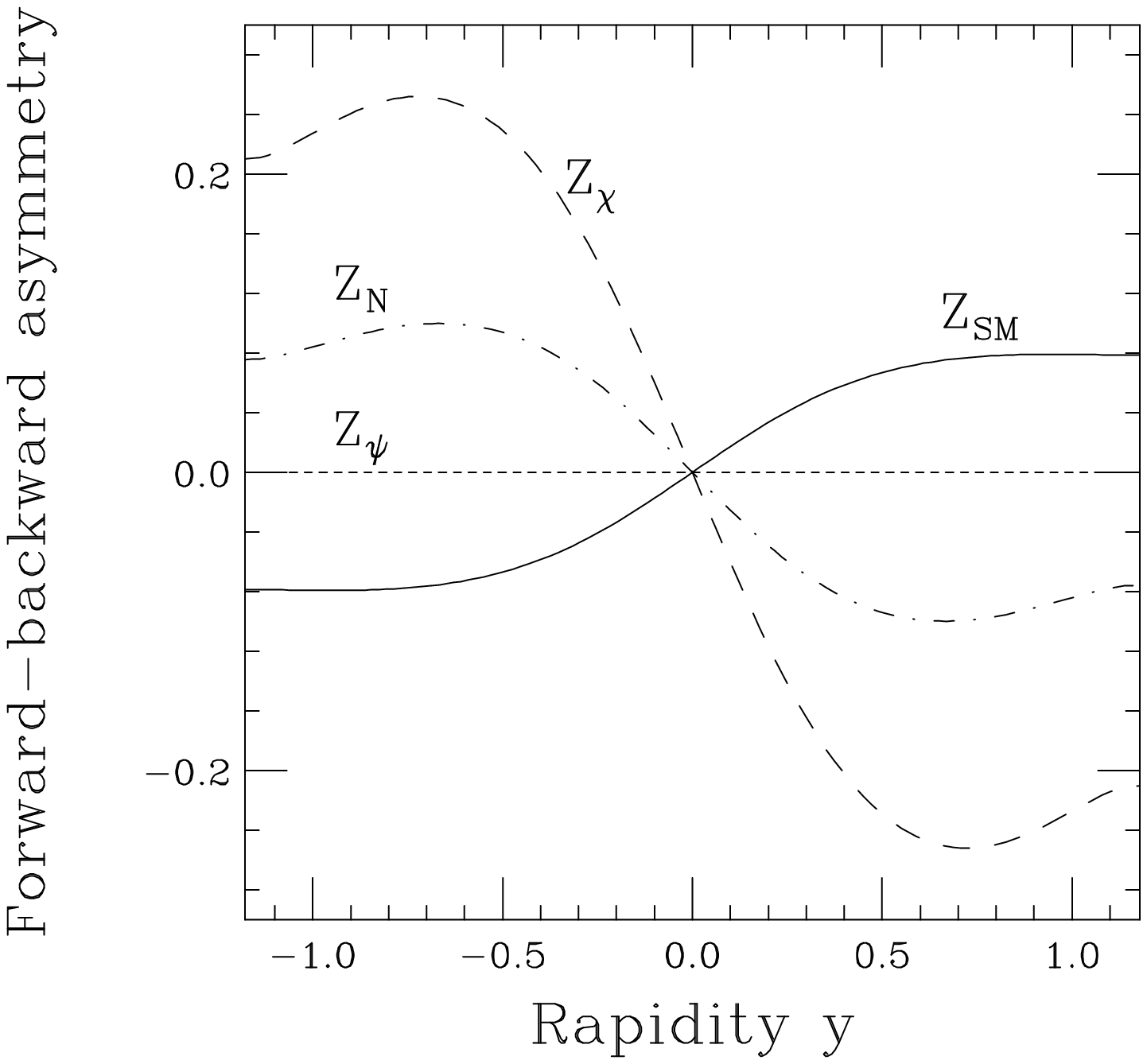}
\end{center}
\caption{Forward-backward asymmetries for several different $Z$s of mass 3
(left) and 4 (right) TeV.
\label{fig:fba}}
\end{figure}

The $Z_N$ is that linear combination $Z_N =-(\sqrt{15} /4)Z_\psi - (1/4)Z_\chi$
[cf.\ Eq.\ (\ref{eqn:qndef})] to which left-handed antineutrinos do not couple.
Consequently, they do not contribute to a triangle anomaly involving the $Z_N$,
and hence are free to acquire large Majorana masses.  The forward-backward
asymmetries for a number of different $Z'$s with masses 3 and 4 TeV at
$\sqrt{s} = 13$ TeV are displayed in Fig.\ \ref{fig:fba}.  The $Z_\psi$ has
purely axial-vector couplings to SM particles, explaining the absence of its
asymmetry.  While the $Z_N$ contains much more $Z_\psi$ than $Z_\chi$,
it does exhibit some asymmetry, about equal in magnitude and opposite in sign
to that of a $Z$ with SM couplings.  The asymmetries for $M(Z') = 4$ TeV have
nearly the same shape as those for $M(Z') = 3$ TeV but in a compressed range of
$y$.

To calculate forward-backward asymmetries $A_{FB}$ in leptonic decays of a $Z$
one needs quark distribution functions $q(x) \equiv xf_q(x)$ and left- and
right-handed couplings to the $Z$ for quarks and leptons. We consider only $u$
and $d$ quarks ($s$ quarks contribute at most a few percent to cross sections
and asymmetries). Let the right-moving proton contribute a parton with momentum
fraction $x_1$, while the left-moving proton contributes a parton with momentum
fraction $x_2$.  These are related to the rapidity $y$ of the parton-parton
CM (which is also the final dilepton CM rapidity) by
\beq
x_1 = \sqrt{\hat s/s}~e^y~,~~x_2 = \sqrt{\hat s/s}~e^{-y}~,
\eeq
where $\hat s$ is the square of the effective mass of the parton-parton
or dilepton system.  (We neglect transverse momenta.)  Then
$A_{FB} = \sigma_{F-B}/ \sigma_{F+B}$, where
\beq
\sigma_{F-B} = C \{ [(u(x_1) \bar u(x_2) - u(x_2) \bar u(x_1)][L_u^2 - R_u^2]
[L_\ell^2 - R_\ell^2] + (u \to d) \} ~,
\eeq
\beq
\sigma_{F+B} = C \{ [(u(x_1) \bar u(x_2) + u(x_2) \bar u(x_1)][L_u^2 + R_u^2]
[L_\ell^2 + R_\ell^2] + (u \to d) \}~,
\eeq
with $C$ a common constant and $L,R$ denoting left- and right-handed couplings.
Only the ratios of these couplings are important, so we quote them without
normalization.  The couplings are shown in Table \ref{tab:lr}; we take $x =
0.2315$.  For the $Z_\psi$, $Z_\chi$, and $Z_N$ we use the charges in Table
\ref{tab:listf}, with suitable sign changes $Q(f_R) = - Q(f^c_L)$.

% This is Figure 8

% This is Table VI
\begin{table}
\caption{Left- and right-handed couplings of $u$ quarks, $d$ quarks, and SM
charged leptons $\ell$ to various $Z$s.
\label{tab:lr}}
\begin{center}
\begin{tabular}{c c c c c c c} \hline \hline
$Z$ type & $L_u$ & $R_u$ & $L_d$ & $R_d$ & $L_\ell$ & $R_\ell$ \\ \hline
 SM & $\frac12 - \frac23x$ & $-\frac23x$ & $-\frac12+\frac13x$ & $\frac13x$ &
 $-\frac12+x$ & $x$ \\
$Z_\chi$ &  --1  &   1   &  --1  &  --3  &    3     &   1 \\
$Z_\psi$ &  1  &   --1   &  1  &   --1   &   1    &   --1 \\
  $Z_N$  &  --1  &   1   &  --1  &   2   &   --2    &   1 \\ \hline \hline
\end{tabular}
\end{center}
\end{table}

A sufficiently heavy $Z_N$ can decay to all the pairs listed in Table
\ref{tab:listf}, diluting its branching fraction to SM particles and eroding
the lower bounds on its mass.  For example, whereas ATLAS places a 95\%
confidence-level lower limit of 4.05 TeV on the mass of a SM $Z'$ based
on 13.3 fb$^{-1}$ at 13 TeV \cite{ATLASZP}, and correspondingly weaker bounds
for $Z_\chi$ and $Z_\psi$, these bounds assume only decays to SM particles.
The same assumption is made by CMS in placing a lower bound of 4.0 TeV on the
mass of $Z'$ \cite{CMSZP}. Suppose all channels in Table \ref{tab:listf} are
open (including right-handed neutrinos).  Then the fraction $r$ of decays to
pairs of the SO(10) 16-plet for a $Z(\theta) \equiv Z_\psi \cos \theta + Z_\chi
\sin \theta$ takes the form $r = (2/9) + (2/9) \sin^2 \theta$, i.e., $(2/9)
\le r \le (4/9)$.  The exotic SO(10) 10-plet and singlet members thus can
affect the reach of $Z'$ searches.  The mass of $Z'$ and a way to escape the
search bounds for the current model are discussed in Sec.~IV.
\bigskip

\centerline{\bf XI.  FUTURE PROJECTIONS}
\bigskip
In the previous sections, we discussed the detection signatures and properties
of the electroweak and TeV scale fermions, scalars, and vectors bosons, as well
as the individual constraints on their masses. The next pertinent question is
how the discovery or non-discovery of any one type of the above will affect the
constraints on the mass spectra of the other types. Another question is whether
such a discovery will exclude the model in some parameter region. In the
following, we quantify these questions for the minimal scenario of three
fermion and one scalar generation of 27-plets of \esix~in the matter sector.

Discovery of a $\gamma\gamma$ resonance will fix the mass and the cross section,
which will fix the value of $\langle \tilde{n}\rangle$ in this model according
to Eq.~(\ref{eqn:vacsigma}). Therefore, Eq.~(\ref{eqn:b17}) would determine
the value of $b_{17}$. Fig.~\ref{fig:cont13} can be used to put bounds on the
allowed Yukawa couplings for the exotic quarks and leptons. 

\begin{figure}
\begin{center}
\includegraphics[width=0.47\textwidth]{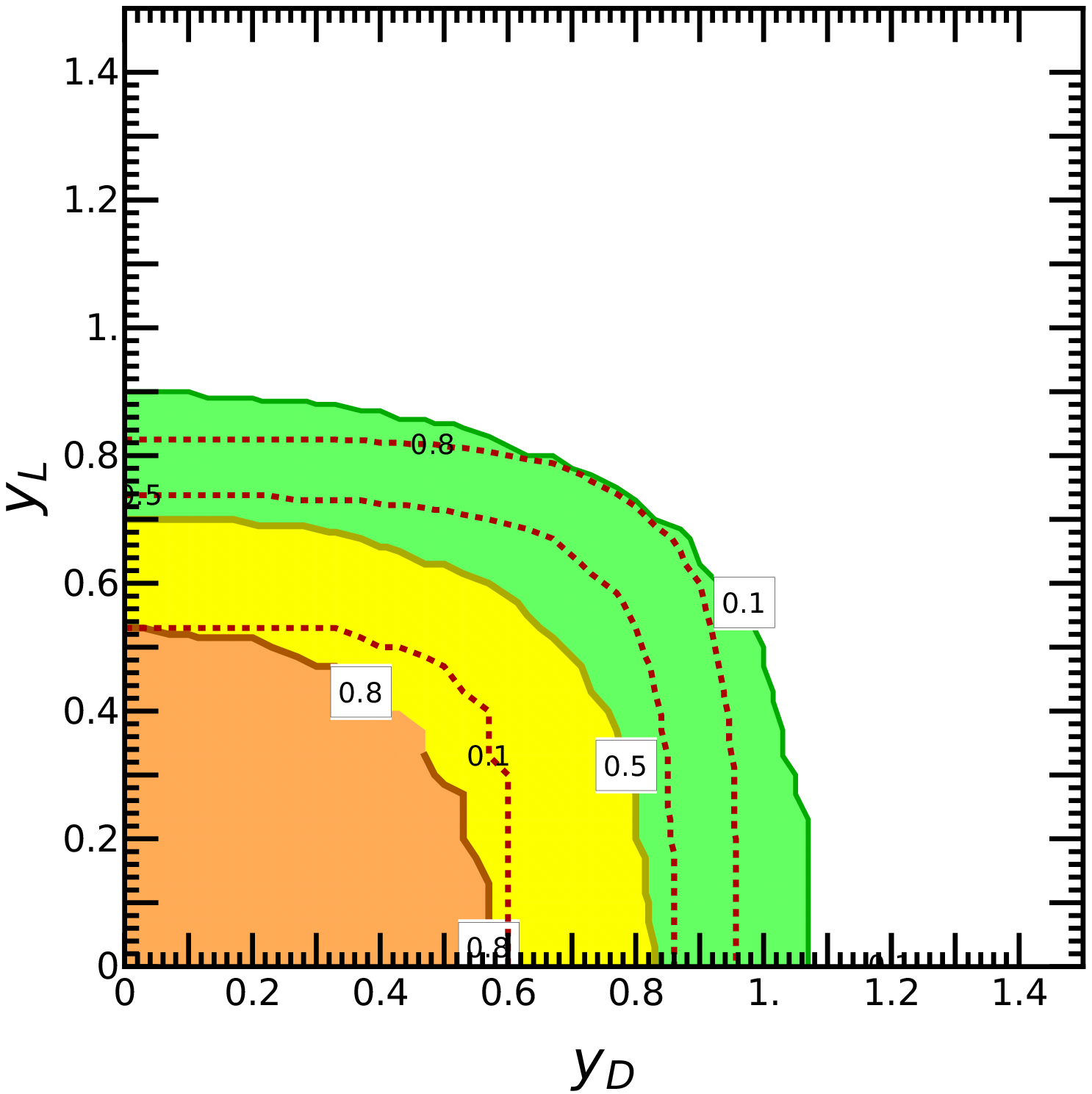}
\includegraphics[width=0.48\textwidth]{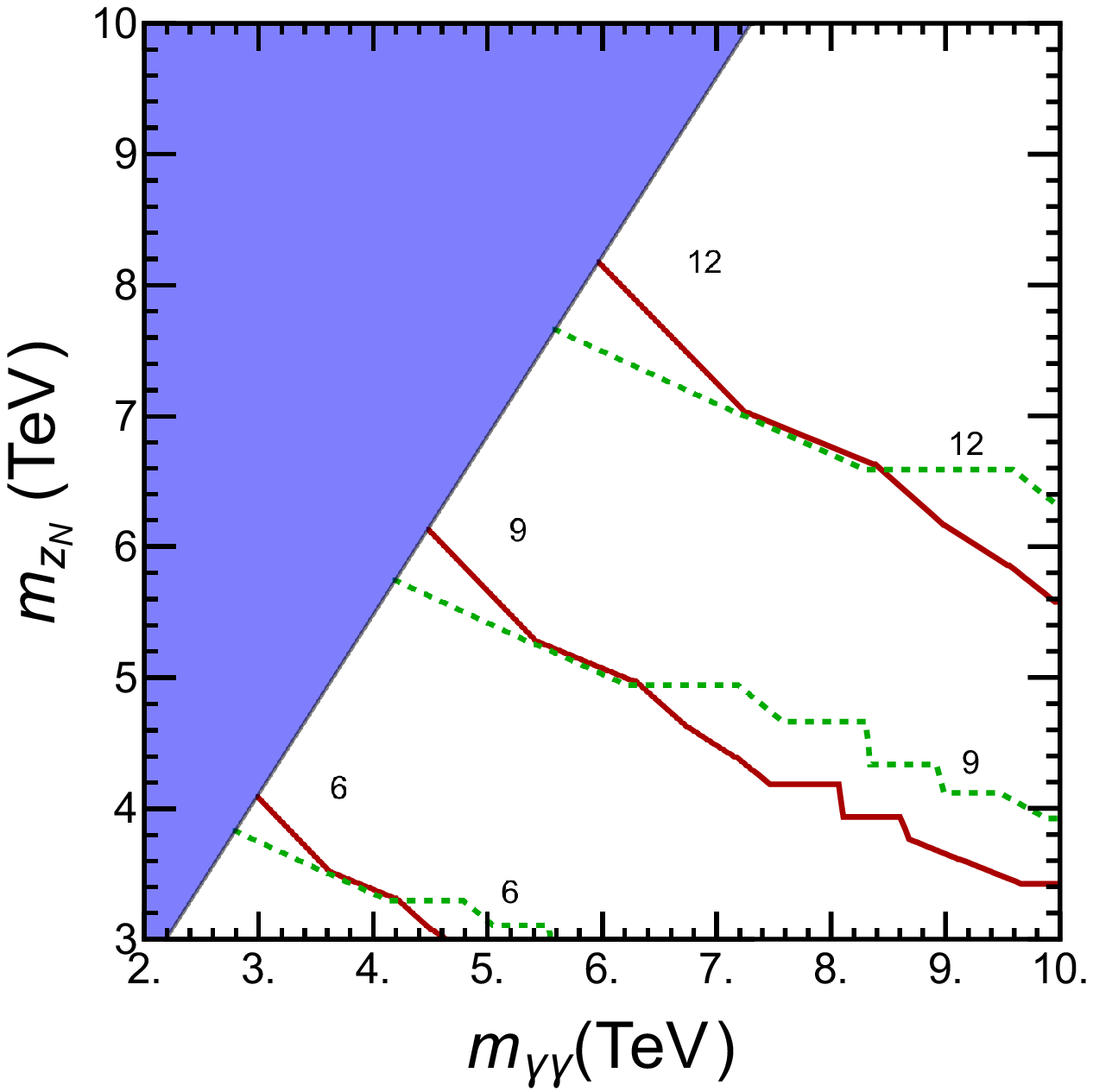}
\end{center}
\caption{\textbf{Left:} Dashed [red] contours correspond to minimum required
value of $b_{17}$ for stability.  Shaded [colored] backgrounds with boundaries
labeled by boxed numbers indicate the total range
of $b_{17}$ values, i.e., the difference between maximum and minimum values of
$b_{17}$ allowed at that point. The inner [orange] region corresponds to the
range of at least $0.8$ above the minimum value given by the [red] dashed
contour.  The middle [yellow] region implies that the allowed $b_{17}$ values
are $0.5$ to $0.8$ more than the minimum values required by the stability
condition as indicated by the [red] dashed contour. The outer [green] region
denotes the corresponding range of 0.1 to 0.5.  \textbf{Right:} Shaded
[light blue] region excluded by stability constraints.  Solid [red] contours
indicate maximum exotic quark mass allowed in TeV. Dashed [green] contours
indicate maximum exotic lepton mass allowed in TeV.}
\label{fig:yhyl}
\end{figure}

In the left panel of Fig.~\ref{fig:yhyl}, the [red] dashed contours show the
minimum value of $b_{17}$ required to avoid an unstable potential. As expected,
it increases with increasing Yukawa coupling because Yukawa contributions $y$
to the RGE of $b_{17}$ are proportional to $-y^4$. After choosing a point in
this plane, the [red] dashed contour value at that point tells the minimum
$b_{17}$ necessary for the stability at that point. In order to calculate the
upper bound of allowed $b_{17}$ at that point, we need the difference between
the upper and lower bound. That difference is given by the [color or] position
of the shaded region. For example, the inner [orange] region corresponds to the
difference between the allowed upper and lower bound of at least $0.8$.
Similarly the middle [yellow] region implies that the allowed values of
$b_{17}$ are $0.5$ to $0.8$ more than the minimum values required by the
stability condition as indicated by the [red] dashed contour. Finally in the
outer [green] region the difference between the allowed upper and lower bound
on $b_{17}$ can vary between $0.1$ to $0.5$. 

Thus discovery of a $\gamma\gamma$ resonance will lead to fixing of $b_{17}$
which can be readily
used to fix the upper and lower bounds on fermion masses using the left panel of Fig.~\ref{fig:yhyl}. Calculations leading to this figure forbid Yukawa
values above 1.3, putting a tight upper bound on the fermion discovery. 

If a $Z_N$ is discovered before the rest of the spectrum, the constraint is
reversed implying a very high value of $\langle \tilde{n}\rangle$, thus fixing
the rest of the mass spectrum at a similar scale. In the right panel of
Fig.~\ref{fig:yhyl}, the boundary of the shaded [light blue] region corresponds
to the minimum value of $b_{17}$ allowed by stability conditions thus excluding
the shaded [light blue] region left of this line. Therefore, a $Z_N$ discovery
will put a lower bound on the mass of the scalar $\tilde{n}$. As $m_{Z_N}$
essentially fixes $\langle\tilde{n}\rangle$, each point in the
$m_{Z_N}-m_{\gamma\gamma}$ space has a fixed value of allowed $b_{17}$, which
can be used to bound the fermion masses using the left panel of
Fig.~\ref{fig:yhyl}. These upper bounds on the fermion masses are shown in the
right panel of Fig.~\ref{fig:yhyl} with [red] solid contours for exotic quarks
and [green] dashed contours for exotic leptons. Contour labels indicate the
masses of fermions in TeV.  Thus the masses of $Z_N$ and $\gamma\gamma$
resonances can fix an upper bound on the fermion masses, leading to exclusion
of the minimal scenario if a heavier fermion is discovered.

% This is Figure 9

If an exotic fermion with mass $M$ is discovered first, that will introduce a
lower bound on the value of $\langle\tilde{n}\rangle$ of about $M$
due to the upper bound on the Yukawa couplings of $~1.3$. This will set a lower
bound on the mass of $Z_N$ and a corresponding lower bound on the mass of an
observble $\gamma\gamma$ resonance. Discovery of either of these below these
bounds will exclude this minimal scenario.  The model can be rescued by adding
another generation of scalars, which has a required SO(10) singlet to give a
mass to the $Z_N$ boson, thus decoupling it from the rest of the constraints.

Forunately, searches at the LHC continue to progress, as exemplified by the
recent ATLAS \cite{ATNOGG,ATZN17,ATVT,ATSQ} and CMS \cite{CMS17001,CMS16008,%
CMS16032,CMS16051} results quoted here.  In the absence of a new discovery,
Fig.~\ref{fig:cont13} also shows the dashed [red] contours with labels
indicating the integrated luminosities at the LHC. The regions to the left of
these type of contours are excluded, if the LHC run up to the indicated
luminosity does not find any $\gamma\gamma$ resonance. The remaining viable
parameter space can be further reduced using the lower bounds on
$\langle\tilde{n}\rangle$ coming from the non-discovery of a $Z_N$ boson and 
exotic fermions.   

\bigskip
%\newpage

\centerline{\bf XII.  CONCLUSIONS}
\bigskip

We have discussed some signatures for TeV-scale physics and for dark matter
candidates of an \esix~scheme. One prediction of this framework is the
existence of a weak isosinglet scalar particle $\tn$ belonging to the same
27-plet as the SM Higgs boson. This picture necessarily requires there to be
a second Higgs boson (not discussed further here) as in all other
two-Higgs-doublet models. The \esix~scheme entails a number of predictions
testable in continued LHC operation at $\sqrt{s} = 13$ TeV. These include
exotic weak isosinglet vector-like quarks ``$D$'' with charges $\mp 1/3$, and
exotic weak isodoublet leptons $L_1^-,~L_1^0,~{L_2^0}^c$, and $L_2^+$.  The
quarks and leptons should have masses at the TeV scale if those masses are
generated by the VEV of $\tn$.  Suggestions for observing them at the LHC
have been made.  There should be a several-TeV-scale $Z_N$ (not coupling to
right-handed neutrinos) whose leptonic decays should exhibit a characteristic
forward-backward asymmetry, odd in rapidity of the dilepton system.

The branching ratios of $\tn$ to $\gamma\gamma,~Z\gamma,~ZZ$, and $WW$ should be
in definite ratios $1:0.24:2.08:5.32$, affording the possibility of early
confirmation or refutation of the model. We also demonstrated how the discovery 
of any of the exotic fermions, bosons or scalar in the minimal scenario of
three fermion generations and a scalar generation can tightly constrain the
rest of the mass spectrum and exclude or strengthen the evidence for an
\esix~scenario based on a subseqent discovery.
\bigskip

\centerline{\bf ACKNOWLEDGMENTS}
\bigskip

We thank Carl Albright, Jun Gao, Sam Harper, Boaz Klima, Joe Lykken, Petra
Merkel, Pavel Nadolsky, Jim Pilcher, Chris Quigg, Michael
Ratz, Tom Rizzo, Lian-Tao Wang, and Carlos Wagner for helpful comments.  This
work was supported in part by the United States Department of Energy through
Grant No.\ DE FG02 13ER41958.  J.L.R. thanks the Mainz Institute
for Theoretical Physics (MITP), the Universit\`a di Napoli Federico II, and
INFN for its hospitality and its partial support during a portion of this work.
MITP is part of the Excellence Cluster PRISMA (EXC 1098) funded by the German
Research Foundation (Deutsche Forschungsgemeinschaft).
%\bigskip
\newpage

\appendix
\centerline{\bf APPENDIX A: Decomposition of 27-plet, 78-plet and $351'$-plet}
\bigskip

The decomposition of the 27-plet, 78-plet and $351'$-plet into the components
invariant under lower-ranked symmetry groups is shown below
\cite{Slansky:1981yr}. $U(1)_\psi$ charges are noted in square brackets,
$U(1)_\chi$ charges in the brackets, and $U(1)_N$ charges in parentheses.

First, $E_6\rightarrow SO(10)\times U(1)_\psi$ breaking leads to the following
decomposition of the $E_6$-invariant multiplets:
\begin{align}
27&=1[4]+10[-2]+16[1]\notag\\
78&=1[0]+16[-3]+\overline{16}[3]+45[0]\notag\\
351'&=1[-8]+10[-2]+\overline{16}[-5]+54[4]+\overline{126}[-2]+144[1]
\end{align}

Next, $SO(10)\times U(1)_\psi\rightarrow SU(5)\times U(1)_\chi\times U(1)_\psi$
breaking leads to the decomposition of the $SO(10)$ invariant multiplets.  The
original $E_6$-invariant multiplets can now be written as the following
$U(1)_\psi\times SU(5)\times U(1)_\chi$ components:
\begin{align}
27&=[4]1\left\{0\right\}+[-2]5\left\{2\right\}+[-2]\overline{5}\left\{-2\right\}
+[1]1\left\{-5\right\}+[1]\overline{5}\left\{3\right\}+[1]10\left\{-1\right\}\notag\\
78&=[0]1\left\{0\right\}+[-3]1\left\{-5\right\}+[-3]\overline{5}\left\{3\right\}
+[-3]10\left\{-1\right\}+[3]\overline{1}\left\{5\right\}+[3]5\left\{-3\right\}
+[3]\overline{10}\left\{1\right\}\notag\\
&+[0]1\left\{0\right\}+[0]10\left\{4\right\}+[0]\overline{10}\left\{-4\right\}
+[0]24\left\{0\right\}\notag\\
351'&=[-8]1\left\{0\right\}+[-2]5\left\{2\right\}+[-2]\overline{5}\left\{-2\right\}
+[-5]\overline{1}\left\{5\right\}+[-5]5\left\{-3\right\}+[-5]\overline{10}
\left\{1\right\}\notag\\
&+[4]15\left\{4\right\}+[4]\overline{15}\left\{-4\right\}+[4]24\left\{0\right\}
\notag\\
&+[-2]\overline{1}\left\{10\right\}+[-2]5\left\{2\right\}+[-2]\overline{10}
\left\{6\right\}+[-2]15\left\{-6\right\}+[-2]\overline{45}\left\{-2\right\}
+[-2]45\left\{2\right\}\notag\\
&+[1]\overline{5}\left\{3\right\}+[1]5\left\{7\right\}+[1]10\left\{-1\right\}
+[1]15\left\{-1\right\}+[1]24\left\{-5\right\}+[1]40\left\{-1\right\}
+[1]\overline{45}\left\{3\right\}
\end{align}

Finally, $SU(5)\times U(1)_\chi\times U(1)_\psi\rightarrow SU(5)\times U(1)_N$
results in the following charge allocations:
\begin{align}
27&=1(-5)+5(2)+\overline{5}(3)+1(0)+\overline{5}(-2)+10(-1)\notag\\
78&=1(0)+1(5)+\overline{5}(3)+10(4)+\overline{1}(-5)+5(-3)+\overline{10}(-4)\notag\\
&+1(0)+10(-1)+\overline{10}(1)+24(0)\notag\\
351'&=1(10)+5(2)+\overline{5}(3)+\overline{1}(5)+5(7)+\overline{10}(6)\notag\\
&+15(-6)+\overline{15}(-4)+24(-5)\notag\\
&+\overline{1}(0)+5(2)+\overline{10}(1)+15(4)+\overline{45}(3)+50(2)\notag\\
&+\overline{5}(-2)+5(-3)+10(-1)+15(-1)+24(0)+40(-1)+\overline{45}(-2)
\end{align}

These results are summarized in Tables \ref{tab:27}--\ref{tab:351c}.
\newpage

% This is Table VII
\begin{table}
\caption{Members of 27-plet of E$_{\rm 6}$, their SO(10) and SU(5)
representations, and their U(1) charges.  The weak hypercharge $Y_W$ is
equal to $2(Q - I_{3L})$.  Mass scale refers to scalar members.  MTeV =
multi-TeV.
\label{tab:27}}
\begin{center}
\begin{tabular}{c c c c c c c c} \hline \hline
   SO(10),& $2\sqrt{6}~Q_\psi$ & $2\sqrt{10}~Q_\chi$ & $2\sqrt{10}~Q_N$ &
 SU(3)$_c$ & SU(2)$_L$ & $Y_W$ & Mass \\
   SU(5) & & & & & & & scale\\ \hline
  16,$\overline{5}$ &   1  &   3  &  --2 & $\overline{3}$ & 1 &  2/3  & GUT \\
           &      &      &      &   1   & 2 & --1   & TeV  \\
   16,10   &      &  --1 &  --1 & $\overline{3}$ & 1 & --4/3 & GUT \\
           &      &      &      &   3   & 2 &  1/3  & GUT \\
           &      &      &      &   1   & 1 &   2   & GUT   \\
    16,1   &      &  --5 &   0  &   1   & 1 &   0   & TeV  \\ \hline
  10,$\overline{5}$ & --2  &  --2 &   3  & $\overline{3}$ & 1 &  2/3  & GUT \\
           &      &      &      &   1   & 2 & --1   & EW/TeV  \\
    10,5   &      &   2  &   2  &   3   & 1 & --2/3 & GUT \\
           &      &      &      &   1   & 2 &   1   & EW/TeV  \\ \hline
    1,1    &   4  &   0  & --5  &   1   & 1 &   0   &  TeV/MTeV  \\ \hline \hline
\end{tabular}
\end{center}
\end{table}
\newpage

% This is Table VIII
\begin{table}
\caption{Members of 78-plet of E$_{\rm 6}$, their SO(10) and SU(5)
representations, and their U(1) charges.  The weak hypercharge $Y_W$ is
equal to $2(Q - I_{3L})$.  Masses are at GUT scale for all scalar members.
SU(5) singlets in 16 and $\overline{16}$ of SO(10) do not get VEVs in order to
preserve the U(1)$_N$ down to multi-TeV scale.
\label{tab:78}}
\begin{center}
\begin{tabular}{c c c c c c c} \hline \hline
SO(10), & $2\sqrt{6}~Q_\psi$ & $2\sqrt{10}~Q_\chi$ & $2\sqrt{10}~Q_N$ &
 SU(3)$_c$ & $SU(2)_L$ & $Y_W$ \\
SU(5) & & & & & & \\ \hline
   45,24   &  0  &  0  &  0  &   8   &  1  &  0  \\  
           &     &     &     &   1   &  3  &  0  \\
           &     &     &     &   3   &  2  &--5/3 \\
           &     &     &     & $\overline{3}$ &  2  & 5/3 \\
           &     &     &     &   1   &  1  &  0  \\
   45,10   &     &  4  & --1 &   3   &  2  & 1/3 \\
           &     &     &     & $\overline{3}$ &  1  &--4/3 \\
           &     &     &     &   1   &  1  &  2  \\
 45,$\overline{10}$ &     & --4 &  1  & $\overline{3}$ &  2  &--1/3\\
           &     &     &     &   3   &  1  & 4/3 \\
           &     &     &     &   1   &  1  & --2 \\
   45,1    &     &  0  &  0  &   1   &  1  &  0  \\ \hline
  16,$\overline{5}$ & --3 &  3  &  3  & $\overline{3}$ & 1 &  2/3  \\
           &     &     &     &   1   & 2 &  --1  \\
   16,10   &     & --1 &  4  & $\overline{3}$ & 1 & --4/3 \\
           &     &     &     &   3   & 2 &  1/3  \\
           &     &     &     &   1   & 1 &   2   \\
    16,1   &     & --5 &  5  &   1   & 1 &   0   \\ \hline
  $\overline{16}$,5 &  3  & --3 & --3 &   3   & 1 & --2/3 \\
           &     &     &     &   1   & 2 &   1   \\
$\overline{16},\overline{10}$ &    &  1  & --4 &   3   & 1 &  4/3  \\
           &     &     &     & $\overline{3}$ & 2 & --1/3 \\
           &     &     &     &   1   & 1 & --2   \\
  $\overline{16}$,1 &     &  5  & --5 &   1   & 1 &   0   \\ \hline
     1,1   &  0  &  0  &  0  &   1   & 1 &   0   \\ \hline \hline
\end{tabular}
\end{center}
\end{table}
\newpage

% This is Table IX
\begin{table}
\caption{Members of $351'$-plet of E$_{\rm 6}$, their SO(10) and SU(5)
representations, and their U(1) charges.  The weak hypercharge $Y_W$ is
equal to $2(Q - I_{3L})$. Masses are at GUT scale for all the members.
\label{tab:351a}}
\begin{center}
\begin{tabular}{c c c c c c c} \hline \hline
   SO(10),& $2\sqrt{6}~Q_\psi$ & $2\sqrt{10}~Q_\chi$ & $2\sqrt{10}~Q_N$ &
 SU(3)$_c$ & SU(2)$_L$ & $Y_W$ \\
   SU(5) & & & & & & \\ \hline
  144,$\overline{45}$   &  1  &  3  & --2 &   8   & 2 &  --1 \\
               &     &     &     &   6   & 1 &  2/3 \\
               &     &     &     &   3   & 2 &  7/3  \\
               &     &     &     &   3   & 1 & --8/3 \\
               &     &     &     & $\overline{3}$ & 3 &  2/3 \\
               &     &     &     & $\overline{3}$ & 1 &  2/3  \\
               &     &     &     &   1   & 2 &  --1  \\
    144,40     &     & --1 & --1 & $\overline{6}$ & 2 &  1/3 \\
               &     &     &     &   8   & 1 &   2  \\
               &     &     &     & $\overline{3}$ & 3 & --4/3 \\
               &     &     &     & $\overline{3}$ & 1 & --4/3 \\
               &     &     &     &   3   & 2 &  1/3  \\
               &     &     &     &   1   & 2 &  --3  \\
    144,24     &     & --5 &  0  &   8   & 1 &   0  \\
               &     &     &     &   3   & 2 & --5/3 \\
               &     &     &     & $\overline{3}$ & 2 &  5/3  \\
               &     &     &     &   1   & 3 &   0  \\
               &     &     &     &   1   & 1 &   0   \\
    144,15     &     & --1 & --1 &   6   & 1 & --4/3 \\
               &     &     &     &   3   & 2 &  1/3  \\
               &     &     &     &   1   & 3 &   2   \\
    144,10     &     & --1 & --1 &   3   & 2 &  1/3  \\
               &     &     &     & $\overline{3}$ & 1 & --4/3 \\
               &     &     &     &   1   & 1 &   2  \\
    144,5      &     &  7  & --3 &   3   & 1 & --2/3 \\
               &     &     &     &   1   & 2 &   1  \\
  144,$\overline{5}$    &     &  3  & --2 & $\overline{3}$ & 1 &  2/3 \\
               &     &     &     &   1   & 2 &  --1 \\ \hline \hline
\end{tabular}
\end{center}
\end{table}
\newpage

%This is Table X
\begin{table}
\caption{Members (contd.) of $351'$-plet of E$_{\rm 6}$, their SO(10) and
SU(5) representations, and their U(1) charges. Masses are at GUT scale for all the members.
\label{tab:351b}}
\begin{center}
\begin{tabular}{c c c c c c c } \hline \hline
   SO(10),& $2\sqrt{6}~Q_\psi$ & $2\sqrt{10}~Q_\chi$ & $2\sqrt{10}~Q_N$ &
 SU(3)$_c$ & SU(2)$_L$ & $Y_W$ \\
   SU(5) & & & & & & \\ \hline
  $\overline{126}$,50   & --2 &  2  &  2  &   8   & 2 &   1  \\
               &     &     &     &   6   & 1 &  8/3  \\
               &     &     &     & $\overline{6}$ & 3 & --2/3 \\
               &     &     &     & $\overline{3}$ & 2 & --7/3 \\
               &     &     &     &   3   & 1 & --2/3 \\
               &     &     &     &   1   & 1 &  --4  \\
$\overline{126}$,$\overline{45}$ &     & --2 &  3  &   8   & 2 &  --1 \\
               &     &     &     &   6   & 1 &  2/3 \\
               &     &     &     &   3   & 2 &  7/3  \\
               &     &     &     &   3   & 1 & --8/3 \\
               &     &     &     & $\overline{3}$ & 3 &  2/3  \\
               &     &     &     & $\overline{3}$ & 1 &  2/3  \\
               &     &     &     &   1   & 2 &  --1  \\
  $\overline{126}$,15   &     & --6 &  4  &   6   & 1 & --4/3\\
               &     &     &     &   3   & 2 &  1/3 \\
               &     &     &     &   1   & 3 &   2   \\
$\overline{126}$,$\overline{10}$ &     &  6  &  1  & $\overline{3}$ & 2 & --1/3 \\
               &     &     &     &   3   & 1 &  4/3  \\
               &     &     &     &   1   & 1 &  --2  \\
   $\overline{126}$,5   &     &  2  &  2  &   3   & 1 & --2/3\\
               &     &     &     &   1   & 2 &   1  \\
   $\overline{126}$,1   &     &  10 &  0  &   1   & 1 &   0  \\ \hline \hline
\end{tabular}
\end{center}
\end{table}
\newpage

%This is Table XI
\begin{table}
\caption{Members (contd.) of the $351'$-plet of \esix, their SO(10) and SU(5)
representations, and their U(1) charges. SU(2) triplet and SU(3) octet are
contained in 54 of SO(10), while diquarks are contained in 54 of SO(10) and
$\overline{16}$ of SO(10) as indicated below with mass scale marked as MTeV. 
SM singlets in (54,24) and $(\overline{16},1)$ cannot get VEVs in order to
preserve $U(1)_N$ symmetry down to the MTeV scale.
\label{tab:351c}}
\begin{center}
\begin{tabular}{c c c c c c c c} \hline \hline
   SO(10),& $2\sqrt{6}~Q_\psi$ & $2\sqrt{10}~Q_\chi$ & $2\sqrt{10}~Q_N$ &
 SU(3)$_c$ & SU(2)$_L$ & $Y_W$ & Mass \\
   SU(5) & & & & & & & scale\\ \hline
    54,15      &  4  &  4  & --6 &   6   & 1 & --4/3 & GUT\\
               &     &     &     &   3   & 2 &  1/3  & MTeV$^a$\\
               &     &     &     &   1   & 3 &   2   & GUT\\
  54,$\overline{15}$    &     & --4 & --4 & $\overline{6}$ & 1 &  4/3  & GUT\\
               &     &     &     & $\overline{3}$ & 2 & --1/3 & GUT\\
               &     &     &     &   1   & 3 &  --2  & GUT\\
    54,24      &     &  0  & --5 &   8   & 1 &   0   & MTeV$^b$\\
               &     &     &     &   3   & 2 & --5/3 & GUT\\
               &     &     &     & $\overline{3}$ & 2 &  5/3  & GUT\\
               &     &     &     &   1   & 3 &   0   & MTeV$^c$\\
               &     &     &     &   1   & 1 &   0   & GUT\\ \hline
  $\overline{16},\overline{10}$  & --5 &  1  &  6  & $\overline{3}$ & 2
 & --1/3 & MTeV$^a$\\
               &     &     &     &   3   & 1 &  4/3  & GUT\\
               &     &     &     &   1   & 1 &  --2  & GUT\\
    $\overline{16}$,5   &     & --3 &  7  &   3   & 1 & --2/3 & GUT\\
               &     &     &     &   1   & 2 &   1   & GUT\\
    $\overline{16}$,1   &     &  5  &  5  &   1   & 1 &   0   & GUT\\ \hline
    10,$\overline{5}$   & --2 & --2 &  3  & $\overline{3}$ & 1 &  2/3  & GUT\\
               &     &     &     &   1   & 2 &  --1  & GUT\\
     10,5      &     &  2  &  2  &   3   & 1 & --2/3 & GUT\\
               &     &     &     &   1   & 2 &   1   & GUT\\ \hline
      1,1      & --8 &  0  & 10  &   1   & 1 &   0   & MTeV$^d$\\ \hline\hline
\end{tabular}
\end{center}
\leftline{$^a$Diquark~~$^b$Color SU(3) octet~~$^c$Weak SU(2) triplet~~
$^d$ SO(10) singlet}
\end{table}
\newpage

\centerline{\bf APPENDIX B: Details of (pseudo)scalar potentials}
\bigskip

For the case of $a=0$ and a new pseudoscalar as the Goldstone boson of
$U(1)_N$, from Eq.~(\ref{eqn:pseudoscal33}) we get
\begin{align}
b_{15} v_d^2 + b_{14} v_u^2 + b_3 \langle\tilde{n}\rangle^2 + m_3^2 &= 0 \\ 
b_{14} v_d^2 + b_{16} v_u^2 + b_4 \langle\tilde{n}\rangle^2 + m_4^2 &=m_A^2 \\ 
b_3 {v_d}^2 + b_4 {v_u}^2 + b_{17} \langle\tilde{n}\rangle^2 + m_5^2 &=0
\end{align}
Substituting these back in Eq.~(\ref{eqn:realscal33}), we get 
\begin{align}
\mathcal{M}_{Ln}^s&=\begin{pmatrix}
2b_{15}v_d^2 & 2b_{14}v_uv_d & 2b_3v_d\langle\tilde{n}\rangle\\
2b_{14}v_uv_d & 2b_{16}v_u^2 + m_A^2 & 2b_4v_u\langle\tilde{n}\rangle\\
2b_3v_d\langle\tilde{n}\rangle & 2b_4v_u\langle\tilde{n}\rangle & 2b_{17}\langle\tilde{n}\rangle^2\\
\end{pmatrix}
\end{align}

This structure implies that the strength of mixing between singlet and doublet
is proportional to $b_3$, $b_4$ and $\langle\tilde{n}\rangle$.  For a scalar
(not pseudoscalar) $\tn$, the small mixing then would imply
\begin{align}
b_{17}\sim \frac{m_{\tn}^2}{2\langle\tilde{n}\rangle^2}.\label{eqn:appA}
\end{align}
For $m_{\tn}\sim{\cal O}$(TeV) this would make $b_{17}\sim 1$, requiring $b_3$
and $b_4$ to be constrained to limit the mixing.  Making the spinless candidate
a pseudoscalar by adding another scalar, this term in the mass matrix becomes
$m_{\tn}^2 +2 b_{17}
\langle\tilde{n}\rangle^2$.  In addition, $b_{17}$ is now free to take on
higher values leading to decoupling of the singlet and relaxation of the
constraints on $b_3$ and $b_4$.
\bigskip

\centerline{\bf APPENDIX C: Details of renormalization group evolution and unification}
\bigskip

\leftline {\bf C.1. Gauge couplings}

The model has $U(1)_Y\times SU(2)_L\times SU(3)_c\times U(1)_N$ symmetry at the
TeV scale before spontaneous breaking. The 1-loop RG equations corresponding to
the gauge coupling constants $(g_i)$ are~\cite{Gross:1973id} 
\begin{align}
\frac{d\alpha_i^{-1}}{dt}&=-\frac{b_i}{2\pi}
\end{align}
where $\alpha_i=\frac{g_i^2}{4\pi}$, $t=\log(\Lambda/GeV)$.

% This is Figure 10
\begin{figure}
\begin{center}
\quad\quad\includegraphics[width=0.98\textwidth]{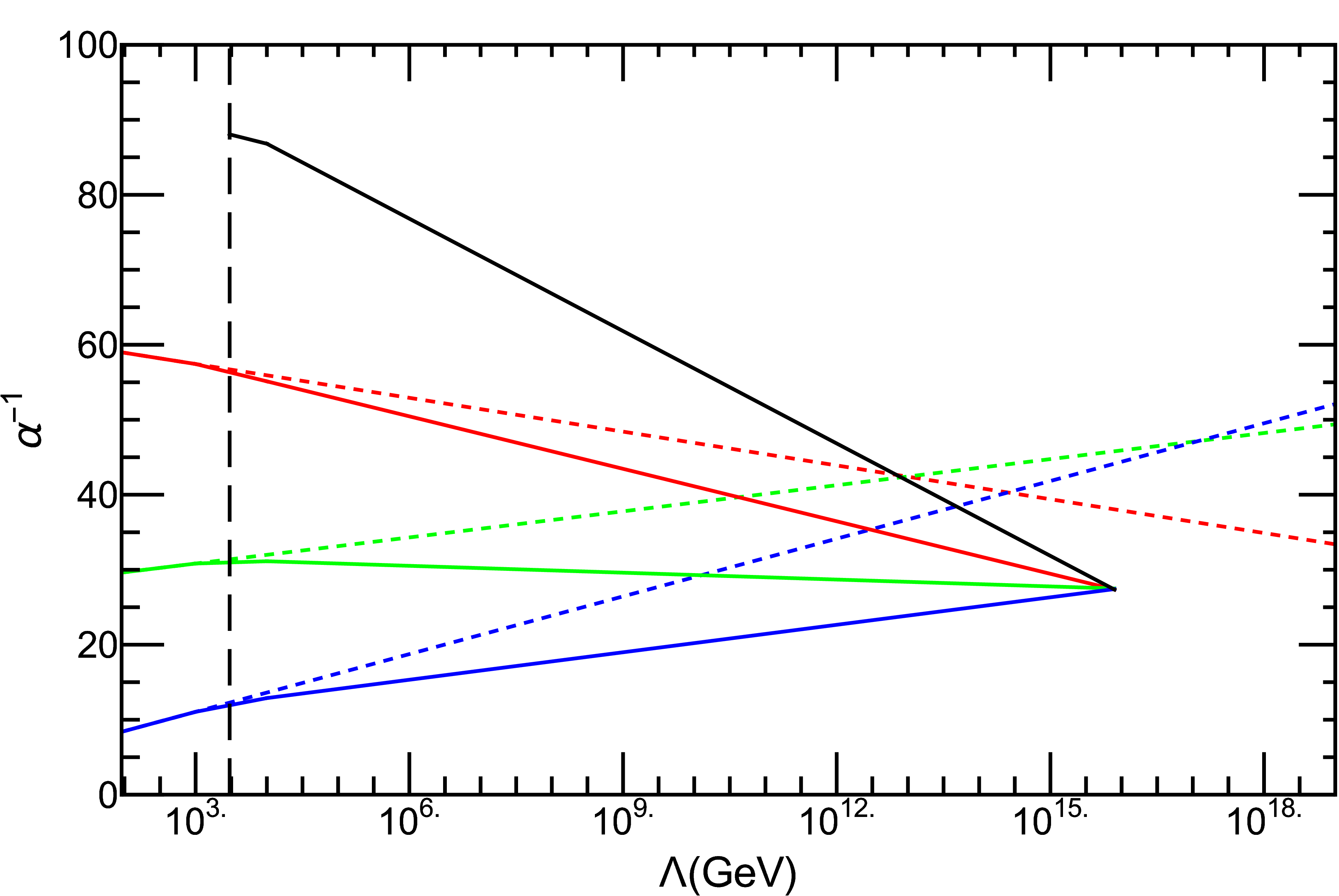}
\end{center}
\caption{RGE for the gauge couplings. \textbf{Dashed}: SM, \textbf{Solid}:
\esix~model with three 27-plet fermions $+$ one 27-plet scalar + one 78-plet
scalar + one $351'$-plet scalar, \textbf{Black} (Upper-most solid):
$\alpha_N^{-1}$ of $U(1)_N$, \textbf{Red} (second from the top for solid,
top-most for dashed at low energies): $\alpha_1^{-1}$ of $U(1)_Y$,
\textbf{Green} (third from the top for solid, second from the top for dashed
at low energies): $\alpha_2^{-1}$ of U(2)$_L$, \textbf{Blue} (third from the
top for solid, third from the top for dashed at low energies): $\alpha_3^{-1}$
of U(3)$_c$. The vertical line corresponds to the experimental lower bound on
the $Z_N$ mass.}
\label{fig:gauge}
\end{figure}

For the SM, we have
\begin{align}
b_1&=\frac{41}{10},\quad b_2=-\frac{19}{6},\quad b_3=-7.%,\quad b_N=\frac{193}{120}
\end{align}

In the \esix~model with $N_f$ generations of fermions and $N_s$ generations of
scalars of 27-plets, we obtain
\begin{align}
b_1&=2\left(N_f+\frac{N_s}{2}\right),\quad b_2=2\left(N_f+\frac{N_s}{2}\right)-\frac{22}{3},\notag\\ b_3&=2\left(N_f+\frac{N_s}{2}\right)-11, \quad b_N=2\left(N_f+\frac{N_s}{2}\right)
\end{align}
The scalar 27-plet generation contributes half of that of a fermion generation,
as it has half the number of degrees of
freedom compared to a Weyl fermion. The above values take into account the full
scalar generation. In order to avoid violation of proton decay bounds, we set
the colored scalar masses $>10^{16}$ GeV for the 27-plet scalar generation. All
of the 78-plet scalar resides at the GUT scale so it does not contribute to the
low-energy evolution of the gauge couplings. For the $351'$-plet scalar, as
described in Sec. II and Appendix A, two copies of diquark, an SU(2) triplet,
an SU(3) octet, and an SO(10) singlet reside at the MTeV scale as indicated in
Table~\ref{tab:351c} of Appendix A.

In making Fig.~\ref{fig:gauge}, for simplicity, we assume that all of the
27-plet exotic fermions and low-energy components of the 27-plet scalar marked
with mass scale other than GUT in Table~\ref{tab:27} are
below or close to $1$ TeV so we include those in the running of the couplings
from $1$ TeV to $10$ TeV. That gives
\begin{align}\label{eqn:neutralbeta}
b_1&=\left(2\,N_f+\frac{3}{10}\right)=\frac{63}{10},\quad b_2=\left(2\,N_f-\frac{22}{3}
+\frac{1}{2}\right)=-\frac{5}{6},\notag\\
b_3&=\left(2\,N_f-11\right)=-5,\quad b_N= \left(2\,N_f + \frac{59}{120}\right)=\frac{779}{120}.
\end{align}

The rest of the low energy states are assumed to reside above $1$ TeV but
below or around $10$ TeV marked as the MTeV mass scale in Table \ref{tab:351c}
in making Fig.~\ref{fig:gauge}. Under the above conditions, the one-loop beta
functions for the evolution from $10$ TeV to the GUT scale are given as
\begin{align}\label{eqn:neutralbeta1}
b_1&=\left(2\,N_f+\frac{3}{10}+2\left(\frac{1}{30}\right)\right)=
\frac{191}{30},\quad b_2=\left(2\,N_f-\frac{22}{3}+\frac{1}{2}+2\left(
\frac{1}{2}\right)+\frac{2}{3}\right)=\frac{5}{6},\notag\\
b_3&=\left(2\,N_f-11+2\left(\frac{1}{3}\right)+1\right)=-\frac{10}{3},\quad
b_N= \left(2\,N_f + \frac{59}{120}+2\left(\frac{9}{5}\right)+\frac{55}{24}
+\frac{5}{6}\right)=\frac{793}{60}.
\end{align}

Figure \ref{fig:gauge} shows the SM gauge RG evolution along with that for the
current model where red (second from the top solid), green (third from the top
solid) and blue (lower-most solid) lines correspond to U(1)$_Y$, SU(2)$_L$, and
SU(3)$_c$, respectively. The black (upper-most solid) line corresponds to
U(1)$_N$ evolution. It is evolved to lower energies starting from the
unification point.  Its intersection with the lower bound on the $Z_N$ mass is
used in Sec.\ IV to derive a lower bound of 6.6 TeV on the VEV of the
SO(10) singlet that breaks the U(1)$_N$ symmetry at the TeV scale.

The unification in Fig.~\ref{fig:gauge} occurs at $8\times10^{15}$ GeV, above
the lower bound on the unification scale
imposed by SuperKamiokande~\cite{Miura:2016krn}. Thus as mentioned before
all the additional particles that could facilitate proton decay can reside at
this energy scale and the symmetry breaking from \esix~down to SM $\otimes$
U(1)$_N$ can occur at this scale via VEVs as described in Sec.\ II and Appendix
A in a manner compatible with experiment. This demonstration of the unification
is only at the one-loop level and with simplifications assumed in the evolution
as mentioned above. Higher-order loops and threshold corrections will alter the
scale at which unification occurs.

At the electroweak scale, the only constraint on the running coupling constant
$g_N$ is that of an upper bound coming from the precise measurement of the Higgs
boson mass which prevents large mixing in the gauge boson mass matrix. If we
change the high energy constraint on $g_N$ in Fig.~\ref{fig:gauge} to demand
unification, that will shift the top-most solid (black) line in the direction
of higher
$\alpha_N^{-1}$ values, thus further reducing the value of $g_N$ at the
electroweak scale which would be allowed by the $Z$ mass constraints. Also,
such an upward shift will always push the Landau pole to higher energies than
the Planck scale Landau pole case. Thus, $g_N$ can always be unified with the
other three coupling constants provided the other three have enough corrections
due to the extra mass above the proton decay scale to enable their unification.
\bigskip

\leftline {\bf C.2. Yukawa and quartic couplings}

The 1-loop renormalization group equations for the Yukawa couplings in this
model can be computed to give the following for the new Yukawa
couplings~\cite{Cheng:1973nv, Ma:1979cw} associated with the vector-like quarks
($y_D$) and ($y_L$):

\begin{align}
16\pi^2\frac{dy_D}{dt}&=y_D\left[\frac{9}{2}y_D^2+2y_L^2-8g_3^2
-\frac{2}{5}g_1^2-\frac{39}{40}g_N^2\right],\\
16\pi^2\frac{dy_L}{dt}&=y_L\left[3y_D^2+\frac{7}{2}y_L^2
-\frac{9}{2}g_2^2-\frac{9}{10}g_1^2-\frac{39}{40}g_N^2\right],
\end{align}
where $t=\ln (\Lambda/GeV)$. The computation can be simplified in the Landau
gauge as noted in~\cite{Cheng:1973nv}. 

\begin{figure}%[Hb]
\begin{center}
\includegraphics[width = 14 cm, clip]{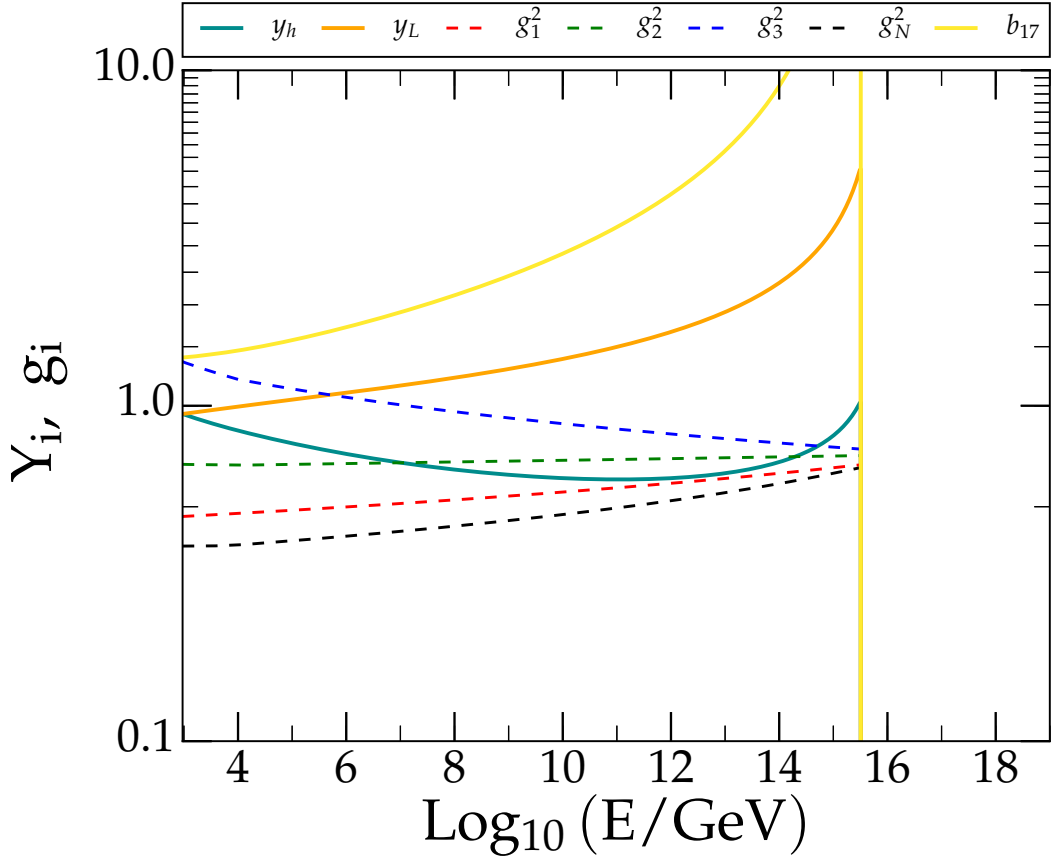}
\end{center}
\caption{RGE for the Yukawa couplings when $g_N$ at MTeV scale is set to be the value required by unification. Dashed curves (upper to lower) correspond to $g_3$ (blue), $g_2$ (green), $g_1$ (red) and $g_N$ (black). Solid curves (upper to lower) correspond to $b_{17}$
(yellow), $y_L$ (orange) and $y_D$ (teal). Electroweak scale Yukawa couplings
are $y_D=0.95$, $y_L=0.95$ and quartic $b_{17}=1.4$.}
\label{fig:yukawa1}
\end{figure}

\begin{figure}
\begin{center}
\includegraphics[width = 14 cm, clip]{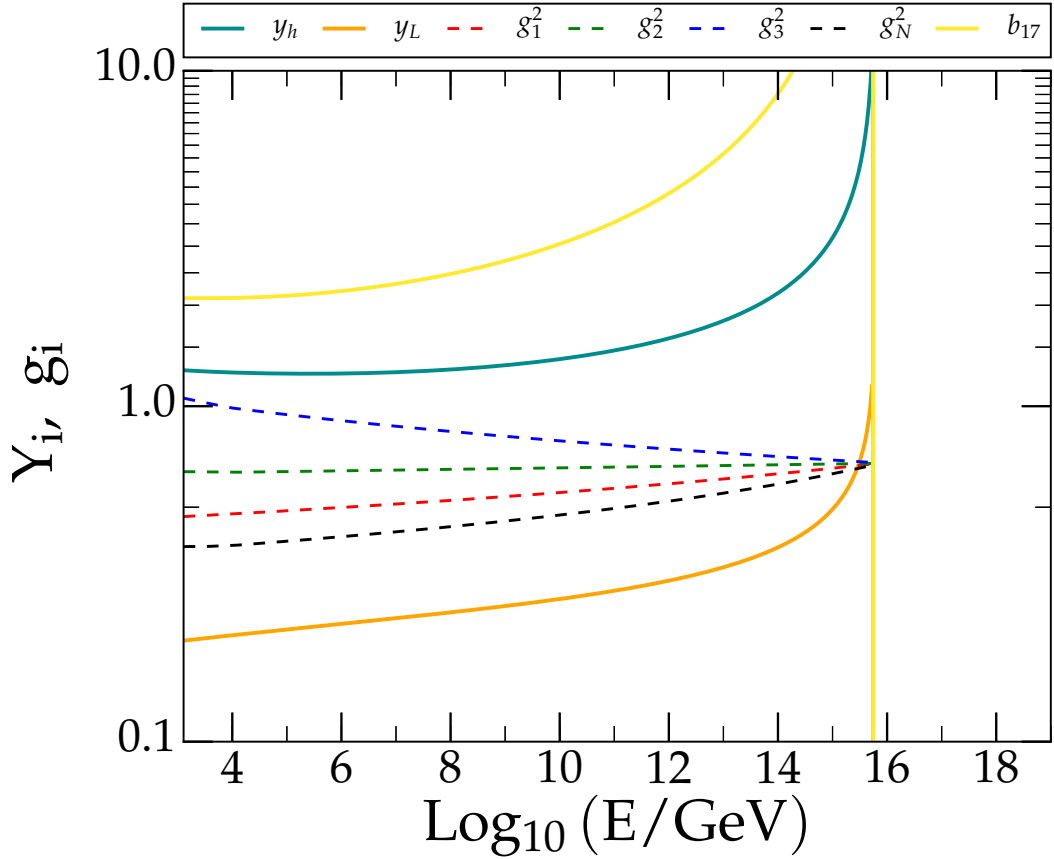}
\end{center}
\caption{RGE for the Yukawa couplings when $g_N$ at MTeV scale is set to be the value required by unification. Dashed curves (upper to lower) correspond to $g_3$ (blue), $g_2$
(green), $g_1$ (red) and $g_N$ (black). Solid curves (upper to lower) correspond to
$b_{17}$ (yellow), $y_D$ (teal) and $y_L$ (orange). Electroweak scale Yukawa
couplings are $y_D=1.3$, $y_L=0.2$, $b_{17}=2.1$.}
\label{fig:yukawa2}
\end{figure}

Figures \ref{fig:yukawa1} and \ref{fig:yukawa2} show the running of the Yukawa
couplings when the
$g_N$ value at $3$~TeV is set to be the one required by the one-loop unification as 
shown in Fig.~\ref{fig:gauge}. Thus this ensures unification as well as the
Landau pole being higher than the Planck scale. The initial condition imposed
for the running of the Yukawa coupling is
$y_i(m_i)=\frac{m_i}{\langle\tilde{n}\rangle}$. For $\langle \tilde n\rangle=1$
TeV, we show the evolution of couplings for $m_D=950$ GeV and $m_L=950$ GeV
with electroweak value of $1.4$ for $b_{17}$ in Fig.~\ref{fig:yukawa1} and for
$m_D =1300$ GeV and $m_L=200$ GeV with electroweak value of $2.1$ for $b_{17}$
in Fig.~\ref{fig:yukawa2}. 

The values of the Yukawa couplings are chosen to demonstrate the maximum
possible values under the condition $y_L=y_D$ (Fig.~\ref{fig:yukawa1}) and maximize the
allowed $y_D$ for the lowest $m_L$ allowed by the experiments (Fig.~\ref{fig:yukawa2}).
The LEP bounds on charged particle searches are $\sim100$ GeV, while the stau
searches at the LHC do not produce bounds stronger than $200$
GeV~\cite{Aad:2014vma,Khachatryan:2014qwa}. Therefore, we are allowed to go as
low as $y_L=0.2$ for $\langle\tilde{n}\rangle=1$ TeV to maximize the allowed
$y_D$. As can be seen, any increase in the masses, i.e., Yukawa coupling values
near the electroweak scale will lead to Landau poles in Yukawa couplings below
$10^{16}$ GeV. 

Similarly reduction in values of the quartic coupling $b_{17}$
will make the potential unstable below the GUT scale, and an increase in the
electroweak scale value of $b_{17}$ will lead to a Landau pole below the GUT
scale. The difference between this maximum value of $b_{17}$ allowed by one loop
perturbativity constraints and minimum value allowed by the stability
constraint increases for smaller values of the Yukawa couplings as shown by the
colored regions in the left panel of Fig.~\ref{fig:yhyl}. 

% This is Figure 11

Finally, we note that owing to the large $Q_N$ charges of $-5/\sqrt{40}$, $\pm
6/\sqrt{40}$ and $10/\sqrt{40}$, the low-energy degrees of freedom belonging to
the $351'$-plet do not contribute to the Yukawa vertex at one-loop level, and
hence do not affect the Yukawa evolution at that level.

% This is Figure 12

\end{document}